\documentclass[preprint,12pt]{elsarticle}

\usepackage{amssymb}
\usepackage{amsmath}

\usepackage{bm}
\usepackage[utf8]{inputenc}
\usepackage[T1]{fontenc}
\usepackage{mathptmx}
\usepackage{etoolbox}
\usepackage{float}

\journal{Chaos, Solitons \& Fractals}

\begin{document}

\begin{frontmatter}



\title{Non-autonomous standard nontwist map}


\affiliation[SP]{organization={Institute of Physics, University of São Paulo},
            addressline={Rua do Matão 1371}, 
            city={São Paulo},
            postcode={05508-090}, 
            state={São Paulo},
            country={Brazil}}

\affiliation[MAR]{organization={Aix-Marseille Université, UMR 7345 CNRS, PIIM},
            addressline={52 Av. Escadrille Normandie Niemen}, 
            city={Marseille},
            postcode={13397 cedex 20}, 
            country={France}}

\author[SP,MAR]{Marcos V. de Moraes} 

\author[SP]{Iberê L. Caldas} 

\author[MAR]{Yves Elskens}

\begin{abstract}
Area-preserving nontwist maps locally violate the twist condition, giving rise to shearless curves. Nontwist systems appear in different physical contexts, such as plasma physics, climate physics, classical mechanics, etc. Generic properties of nontwist maps are captured by the standard nontwist map, which depends on a convection parameter $a$ and a modulation coefficient $b$. In the spirit  of non-autonomous systems, we consider the standard nontwist map (SNM) with a linearly increasing modulation coefficient, and we investigate the evolution of an ensemble of points on the phase space that initially lies on the shearless invariant curve in the initial state, called shearless snapshot torus. Differently from the SNM with constant parameters --- where we can see different scenarios of collision/annihilation of periodic orbits leading to global transport, depending on the region in the parameter space --- for the SNM with time-dependent parameters, the route to chaos is not only related to the path in the $(a, b)$ parameter space, but also to the scenario of the evolution of parameter $b_n$. In this work, we identify power-law relationships between key parameters for the chaotic transition and the iteration time. Additionally, we analyze system reversibility during the chaotic transition and demonstrate an extra transport, where parameter variation modifies the diffusion coefficient.
\end{abstract}

%


\end{frontmatter}



\section{Introduction}
\label{sec1}

In Hamiltonian systems, trajectories in phase space evolve under the influence of Hamilton's equations, which describe the dynamics of canonical coordinates and momenta \cite{Lichtenberg_AJ:2ed:Regular_and_chaotic_dynamics}. Hamiltonian dynamics extends beyond familiar systems, not only well-known conservative ones such as the pendulum \cite{Lowenstein_JH:2012:Essentials_of_Hamiltonian_Dynamics} but also more complex applications, for example, in the study of transport in plasma physics, fluid dynamics, condensed matter, celestial mechanics and other areas \cite{Portela_JSE:2007:Diffusive_transport_throught_a_nontwist_barrier_in_tokamaks, Hazeltine_JD:2003:Plasma_Confinement, Ottino_JM:1989:The_Kinematics_of_Mixing, Pierrehumbert_RT:1991:Large-scale_horizontal_mixing_in_planetary_atmospheres, Kyner_WT:1973:Invariant_manifolds_in_celestial_mechanics, Sousa_MC:Energy_exchange_coupled_system}. 

The dynamics of a time-independent Hamiltonian system with $N$ degrees of freedom can be visualized in the $2N$-dimensional phase space of coordinates and momenta. From an initial value of coordinates and momenta, one can study the dynamics by looking at the trajectory evolution in the phase space. However, finding trajectories in the phase space may be very difficult, as Hamilton`s equations are usually not analytically solvable, and numerical solutions might be quite CPU-consuming. One way to simplify the issue is to examine the Poincaré section of the system. For instance, a three-dimensional Hamiltonian system can be analyzed by observing the values of two variables at the moment the third variable reaches a specific value. The Poincaré section can be represented by discrete-time equations called maps, that can arise naturally in Hamiltonian systems \cite{Lichtenberg_AJ:2ed:Regular_and_chaotic_dynamics}.  

In the context of Hamiltonian dynamics, non-autonomous systems represent a class of dynamical systems where the governing equations of motion explicitly depend on time. Unlike autonomous systems, where the phase space shows usually the coexistence of regular and chaotic trajectories, non-autonomous Hamiltonian systems exhibit time-varying behaviors that depend on the range of the dynamical parameter and the parameter's evolution scenario \cite{Jánosi_D:2019:Chaos_In_hamiltonian_systems_subjected_to_parameter_drift, Jánosi_D:2021:Chaos_in_conservative_discrete-time_systems_subjected_to_parameter_drift}. 
These systems find widespread application across diverse fields, including celestial mechanics, plasma physics, and climate sciences. 

While nontwist maps have been widely studied \cite{Negrete_DDC:1996:Area_preserving_nontwist_maps, Negrete_DDC:1997:Renormalization_and_transition_to_chaos_in_area_preserving_nontwist_maps}, the bibliography for time-varying effects in them is still scarce. In that sense, non-autonomous Hamiltonian systems pose unique theoretical challenges and opportunities, requiring novel techniques to describe their behavior. For instance, the time dependence in these systems breaks conserved quantities and invariant structures, such as invariant tori and periodic orbits, which are central to understanding the system's long-term behavior. Moreover, understanding the chaotic transition is central to the problem since theorems such as KAM do not apply to time-varying maps. 

In this paper, we explore the behavior of the shearless transport barrier --- a distinctive feature of nontwist systems --- under the influence of a time-varying parameter in the standard nontwist map (SNM). Our key findings show that for small variations in this parameter, the evolution of the shearless ensemble closely follows the stationary shearless curve. However, after certain bifurcations, such as separatrix reconnection, the ensemble diverges from the stationary curve. The chaotic transition of the shearless ensemble is characterized by three critical values: the parameters \( b_c^{-} \) (marking the start) and \( b_c^{+} \) (marking the end) of the chaotic transition, as well as the instantaneous Lyapunov exponent, measured during the transition. We provide numerical results for these values as a function of the parameter's evolution. Additionally, we qualitatively track the chaotic transition by evaluating the reversibility of the shearless ensemble throughout the process. Finally, we demonstrate the occurrence of an extra diffusion linked to the varying parameter.

The structure of the paper is as follows: Section II reviews key properties of the SNM. Section III introduces the non-autonomous SNM, incorporating the time-dependent parameter. Section IV investigates the system's behavior under small time-dependent perturbations. Section V focuses on the transition to chaos in the shearless snapshot torus. Section VI briefly addresses the transport properties of the non-autonomous SNM. Finally, Section VII presents our conclusions.

\section{Standard nontwist map}

The standard nontwist map (SNM) is a paradigmatic area-preserving map that locally violates the twist condition (cf. Eq. \eqref{eq:twist_condition}). The SNM $M_0$ reads \cite{Negrete_DDC:1992:Chaotic_Transport_By_Rosby_Waves_In_Shear_Flow}
\begin{equation}
  M_0 = \left\{
    \begin{aligned}
      &     y_{n+1} = y_n - b \sin(2\pi x_n) \\
      &     x_{n+1} = x_n + a (1 - y_{n + 1}^2) 
    \end{aligned}
  \right.
  \label{map-SNM}
\end{equation}
where $a \in [0, 1)$ and $b \in \mathbb{R}$, and the domain of the variables is $D :=  \{(x, y)\ |\ y \in (-\infty, \infty),\ x \in [-1/2, 1/2),\ \ \operatorname{mod} 1 \}$. Variable $x$ plays the role of an angle, and we call $y$ a radius for simplicity. The index $n$ represents a discrete time, which is called iteration from here on. The parameter $a$ shapes the rotation number along the $y$ direction, therefore called convection coefficient. The parameter $b$ represents the amplitude of a radial perturbation, and we called it modulation coefficient. 

In general, we can define the rotation number $\omega$ of an orbit initiated at the point $(x_0, y_0)$, when it exists, by 
\begin{equation}
    \omega(x_0, y_0) = \lim_{n \to \infty} \frac{x_n}{n}, 
    \label{eq:rot_number}
\end{equation}
where the $x$ variable is lifted to the real numbers. 

In particular, if an orbit is periodic, then its rotation number is rational, written as the ratio of two integers $\omega = m/n$. Trajectories with irrational rotation numbers populate densely one-dimensional lines called invariant tori, or more complex objects such as cantori \cite{Meiss92}.

\begin{figure}[H]
    \centering
    \includegraphics[height = 5cm]{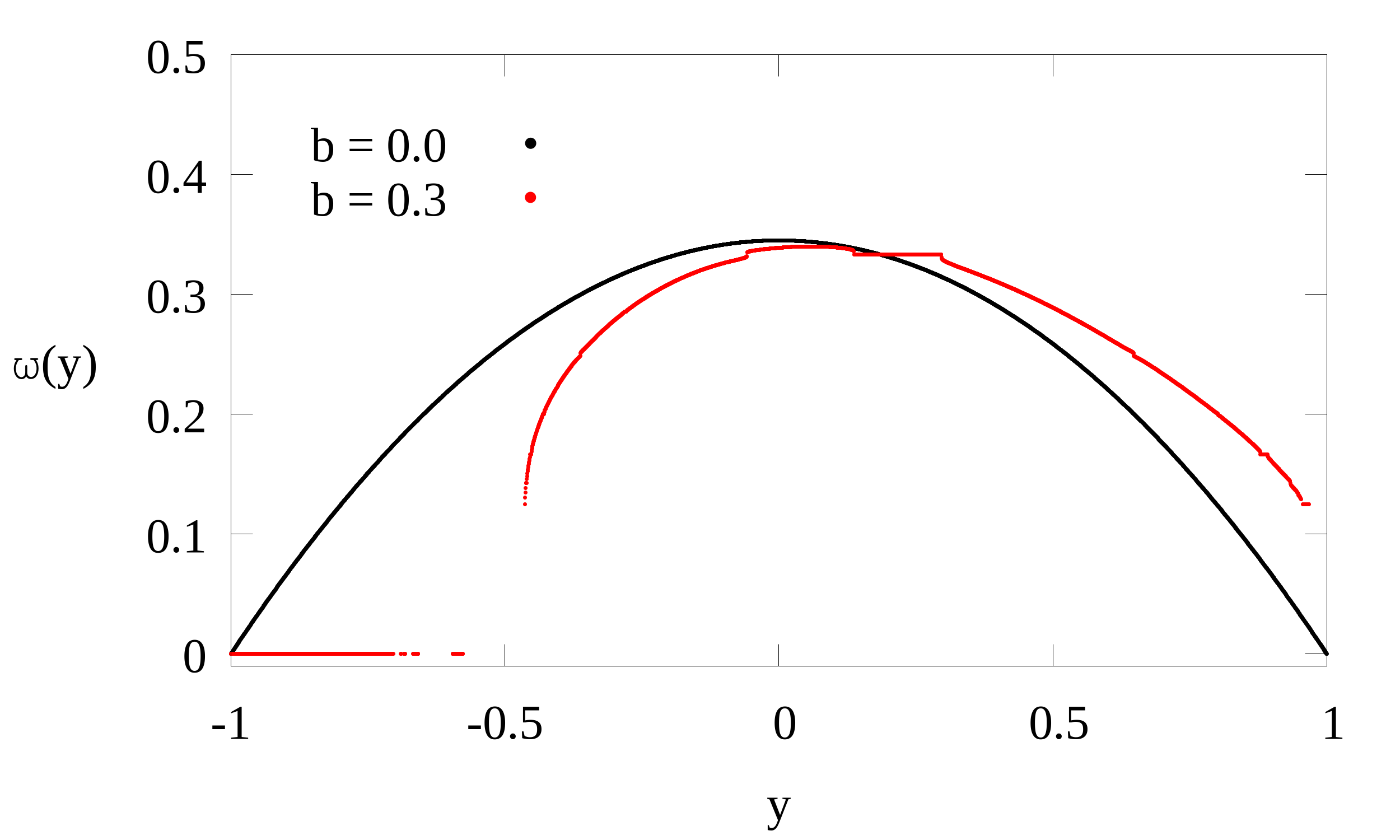}  
    \caption{Rotation number of SNM in function of $y$ coordinate for $b=0$ (black) and $b=0.3$ (red). We consider parameter $a=0.345$. }
    \label{rotation_number}
\end{figure}

In the integrable limit, $b = 0$, successive iterations of an orbit result in a straight line that wraps around the $ x-$domain. For $b \neq 0$, some invariant curves are destroyed, resulting in a mixed phase space with chaotic regions, while other invariant tori still exist. Figure \ref{rotation_number} shows the rotation number for both integrable and non-integrable cases.  Because the rotation number is nonmonotonic, the twist condition
\begin{equation}
    \partial \omega / \partial y \neq 0 
    \label{eq:twist_condition}
\end{equation}
is violated, giving rise to the shearless curve. 

As the KAM theorem assumes the twist condition, the problem of motion stability in nontwist systems involves understanding how the shearless curve --- the curve where the KAM theorem is not valid—breaks up (see \cite{Negrete_DDC:1996:Area_preserving_nontwist_maps, Negrete_DDC:1997:Renormalization_and_transition_to_chaos_in_area_preserving_nontwist_maps}).

Because the map can be factored as a product of involutions, we can determine indicator points (IPs); if these points belong to a regular orbit, then the orbit is the shearless curve, and the IPs are located on it. The indicator points for the SNM are \cite{Shinohara_S:1998:Indicators_of_reconnection_processes_and_transition_to_global_chaos_in_nontwist_maps}

\begin{equation}
    \mathbf{z}_0^{(\pm)} = (\pm 1/4, \pm b/2),\ \ \ \mathbf{z}_1^{(\pm)} = (a/2 \pm 1/4, 0).\ \ \
    \label{eq:IP}
\end{equation}

\begin{figure}[H]
    \centering
    \includegraphics[height = 4cm]{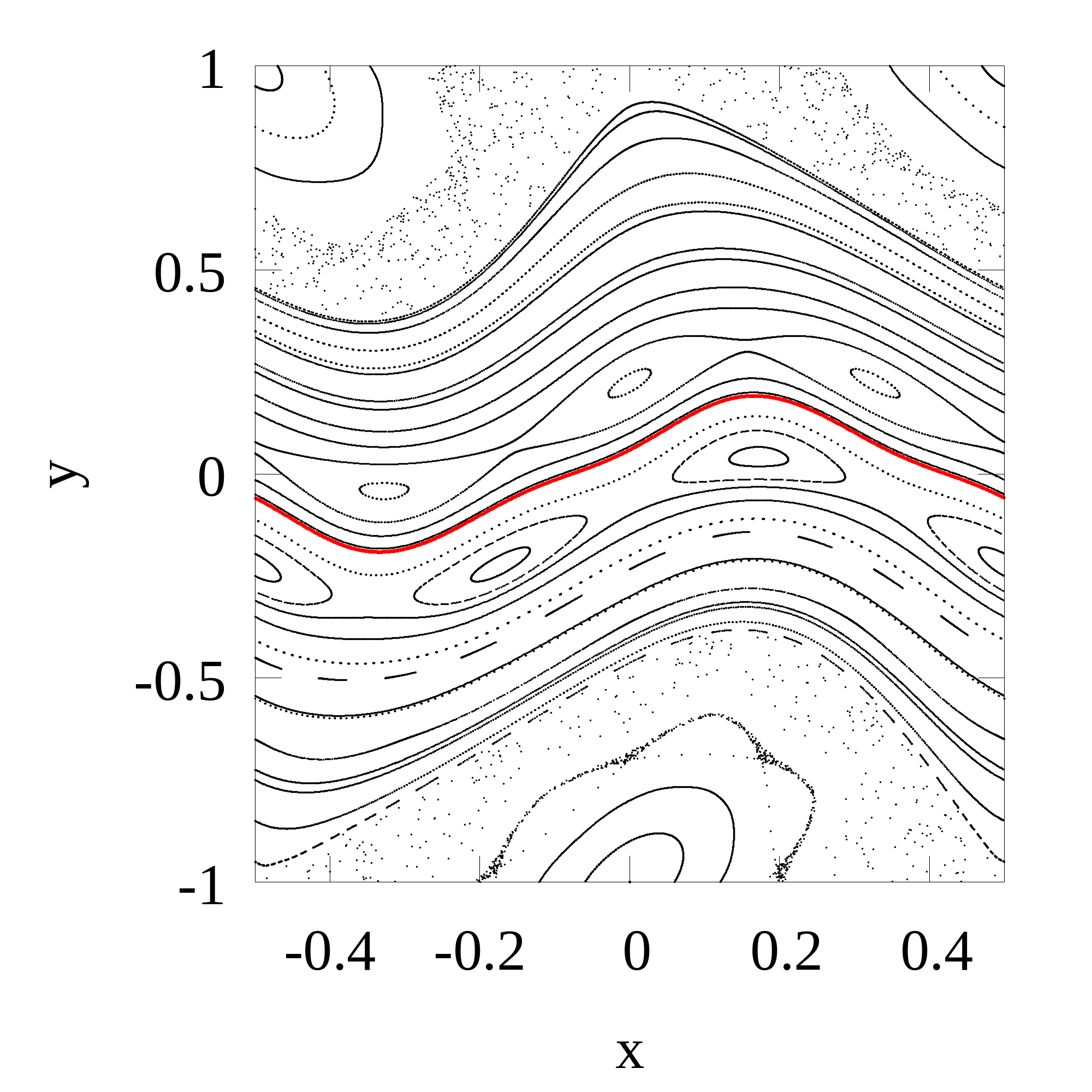}   
    \includegraphics[height = 4cm]{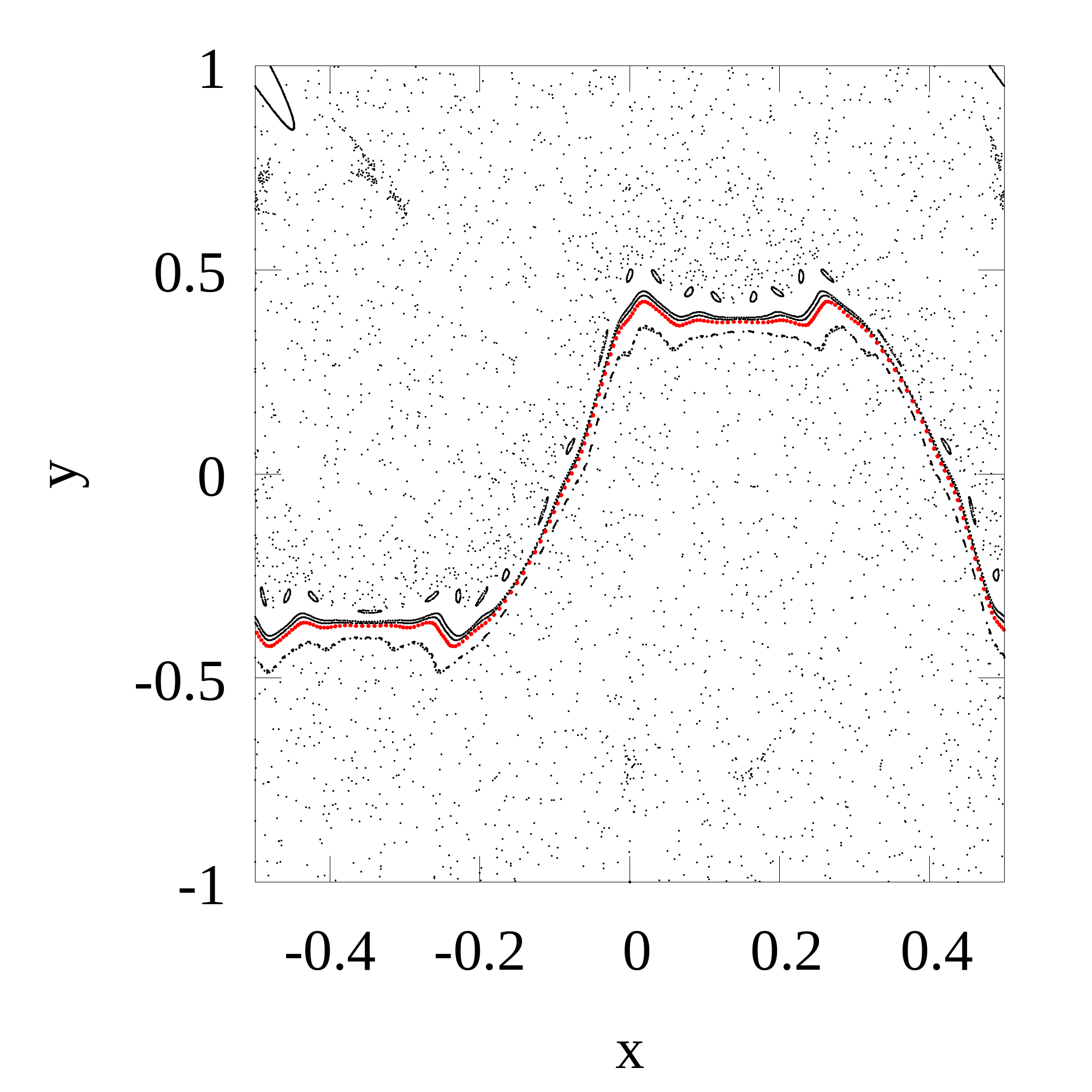}   
    \caption{Phase portraits of the SNM for fixed $a = 0.345$. On the left $b=0.3$, and $b=0.79$ on the right. The shearless curve is plotted in red.}
    \label{phase_portraits}
\end{figure}

Figure \ref{phase_portraits} shows typical phase portraits of the SNM for fixed $a=0.345$. 

\begin{figure}[H]
    \centering
    \includegraphics[height = 4cm]{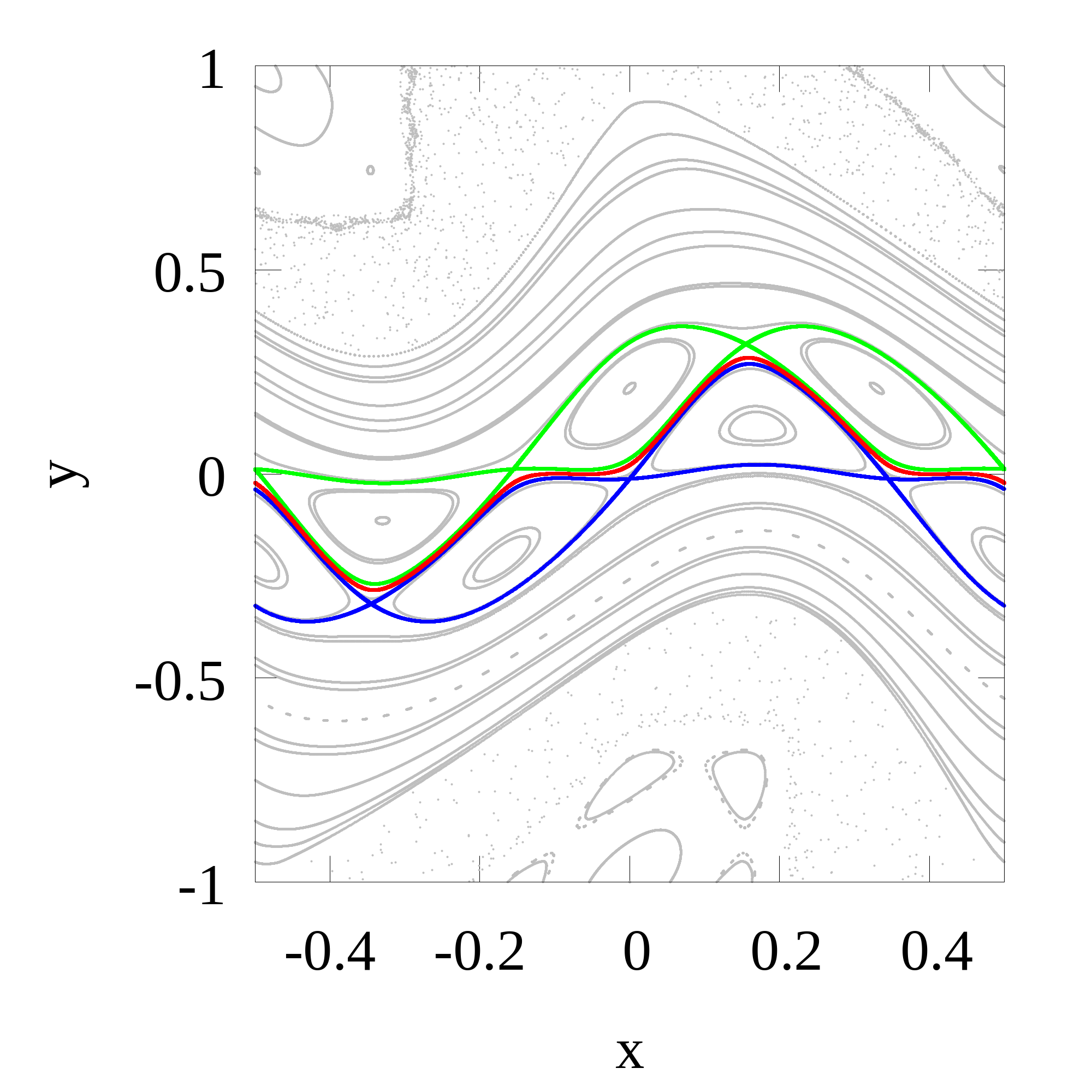}
    \includegraphics[height = 4cm]{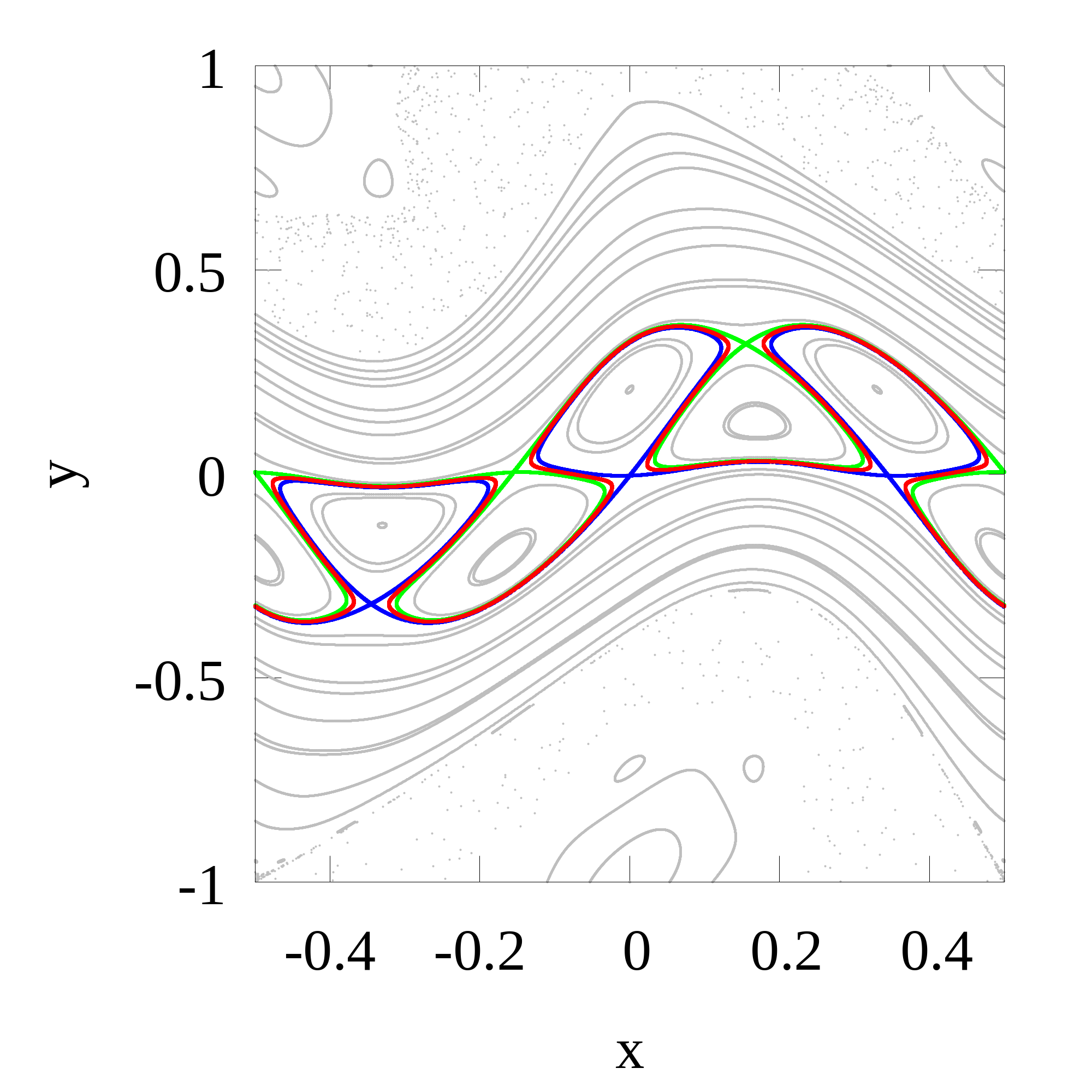}
    \caption{Phase portraits for the SNM with constant parameters, showing in green (blue) the up (down) separatrix and in red the shearless curve. On the left side $b=0.37$, and on the right $b=0.38$.}
    \label{separatrix_reconnection}
\end{figure}

Because of the twist condition violation, periodic orbits with the same rotation number come in pairs. For the integrable case with $a = 0.345$, two period-3 elliptic orbits are present in the phase space. When $b \neq 0$, the system becomes non-integrable, and the period-3 elliptic orbits transform into island chains separated by a shearless barrier. Additionally, two period-3 hyperbolic points exist in this region, each with its own separatrix. These separatrices eventually collide and reconnect, undergoing a global bifurcation known as separatrix reconnection, which changes the phase space topology from heteroclinic to homoclinic near the shearless curve (see Figure \ref{separatrix_reconnection}). The separatrix reconnection threshold for period-one orbits may be found analytically, considering that the hyperbolic point in each separatrix will have the same value of the Hamiltonian \cite{Negrete_DDC:1996:Area_preserving_nontwist_maps}. For higher-order periodic orbits (\( n \geq 2 \)), analytical expressions for the reconnection threshold are unavailable. However, our numerical analysis shows that, for \( a = 0.345 \), separatrix reconnection occurs at \( b^* = 0.3761 \).

For sufficiently high values of the modulation coefficient $b$ (right panel on figure \ref{phase_portraits}) almost all invariant curves are destroyed, with the shearless usually being the last one. The critical value of $b$ for shearless torus breakup is, for $a=0.345$, $b_c \approx 0.83$. After this value the shearless torus is destroyed, but there might be still some partial barrier effects \cite{Szezech_JD:2009:Transport_properties_in_nontwist_area-preserving_maps, Szezech_JD:2012:Effective_transport_barriers_in_nontwist_systems}. 

\section{Parameter drift}

This section introduces the non-autonomous standard nontwist map (NASNM). For that, we consider the parameter $b$ in equation \eqref{map-SNM} to be time-dependent
\begin{equation}
    b \rightarrow b_n. 
\end{equation}

There are several ways to choose the time dependence. For simplicity, we choose
\begin{equation}
    b_n = b_0 + nv,
\end{equation}
where $b_0$ is the initial parameter, $n$ is the iteration number and $v$ is the parameter growth rate, written as $v = (b_f - b_0)/N$. The last iteration satisfies the condition $b_N = b_f$, where $b_f$ is the final parameter and $N$ is the total number of iterations. We write $[b_0, b_f, N]$ to denote the drift scenario.

Discrete-time conservative systems are usually considered under the assumption of constant parameters. When parameter drift is introduced, invariant structures like KAM tori are lost, making the analysis of individual trajectories less meaningful. Instead, a more effective approach is to track ensembles that initially correspond to the KAM curves of the original system. \cite{Jánosi_D:2019:Chaos_In_hamiltonian_systems_subjected_to_parameter_drift, Jánosi_D:2021:Chaos_in_conservative_discrete-time_systems_subjected_to_parameter_drift}. We define an initial ensemble as a sequence of points ${\mathbf{r}}_i$ on the phase space lying in an invariant curve
\begin{equation}
    {\mathbf{R}}_0 = \{\mathbf{r}_0, \mathbf{r}_1,\mathbf{r}_2, \dots, \mathbf{r}_i\}. 
\end{equation}
\sloppy Considering the shearless invariant torus, for the given fixed $b_0$, we choose one of the indicator points $\mathbf{z}_0,\mathbf{z}_1$ as the initial point ${\mathbf{r}}_0$ and the successive points $\{\mathbf{r}_1,\mathbf{r}_2, \dots, \mathbf{r}_i\}$ are obtained from the iteration of ${\mathbf{r}}_0$ with the SNM with fixed parameter $b_0$. The temporal evolution of the ensemble ${\mathbf{R}}_0$ is given by the successive application of the NASNM. At each iteration denoted by $n$, the current image of this ensemble is called snapshot torus. It is essential to distinguish that the snapshot torus, rather than an invariant curve, undergoes this evolution. 

\begin{figure}[H]
    \centering
    \includegraphics[height = 4cm]{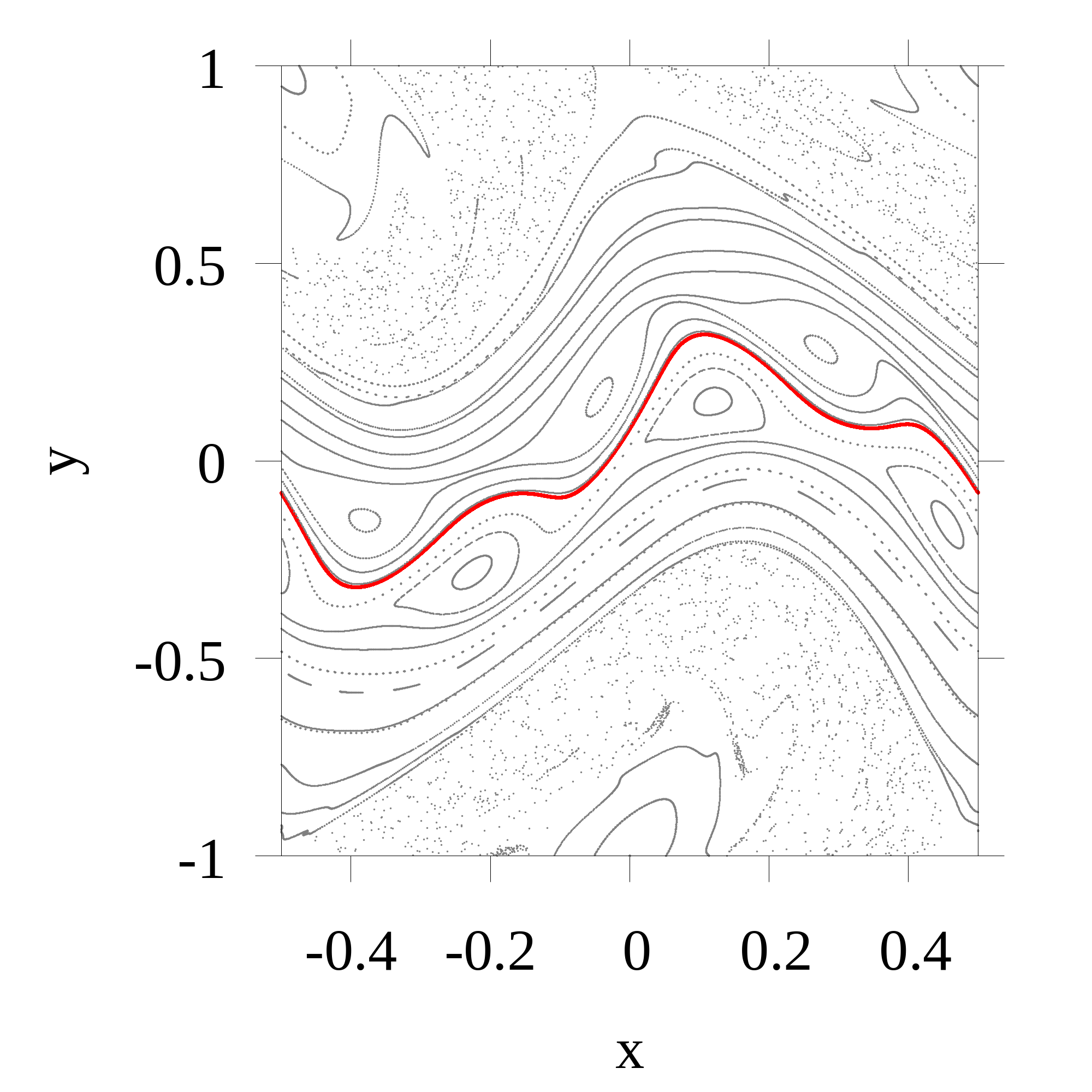}
    \includegraphics[height = 4cm]{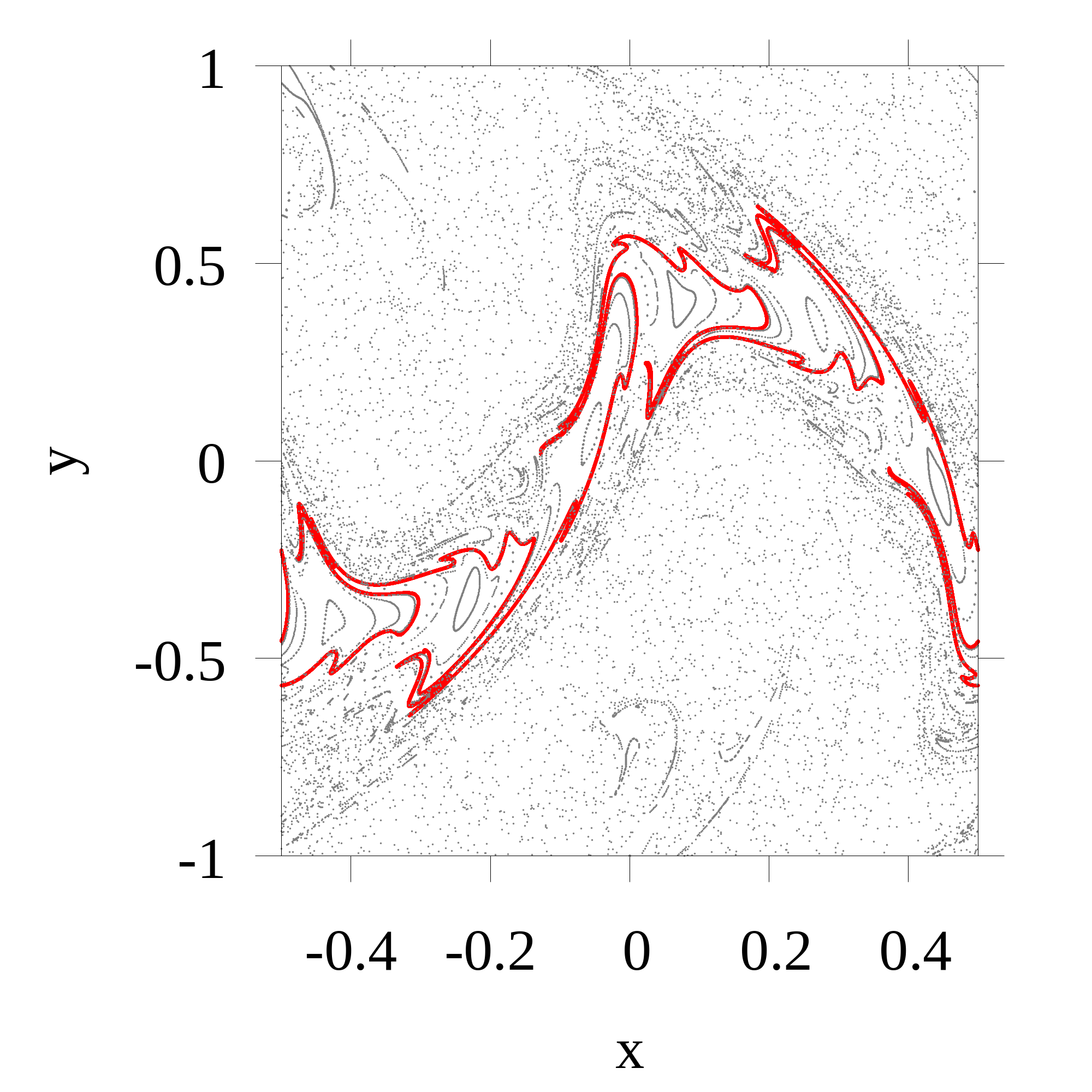}
    \caption{Snapshots of the non-autonomous SNM for $a=0.345$ and drift scenario $[b_0=0.3, b_f= 1.0, N = 100]$, corresponding to the instants $n = 20$ $(b_n = 0.44)$ on the left, and $n = 70$ $(b_n = 0.79)$ on the right. The shearless snapshot torus is plotted in red.}
    \label{snapshot_1}
\end{figure}

Figure \ref{snapshot_1} displays a typical snapshot tori evolution, corresponding to the drift scenario $[b_0=0.3, b_f= 1.0, N = 100]$. It is clear that introducing time dependence significantly changes the phase portrait. During earlier time intervals, expressly when $b_n = 0.44$, the system still retains some characteristics of the SNM, evidenced by the resemblance of the snapshot tori to the invariant curves observed in the stationary model. However, as the system evolves, for instance, when $b_n = 0.79$, the phase portrait deviates substantially from the stationary model. Notably, while the stationary system has already lost all invariant curves --- the shearless invariant curve usually being the last one ---, on the time-dependent model the shearless snapshot torus exhibits a different structure, that is affected by various phenomena observed in the stationary model, such as separatrix reconnection, the emergence of small islands, the destruction of primary islands, and the formation of meanders. In this sense, the ensemble preserves a memory of the evolution scenario. 

\section{Adiabatic variation and separatrix reconnection}

\begin{figure}[H]
    \centering
    \includegraphics[height = 4cm]{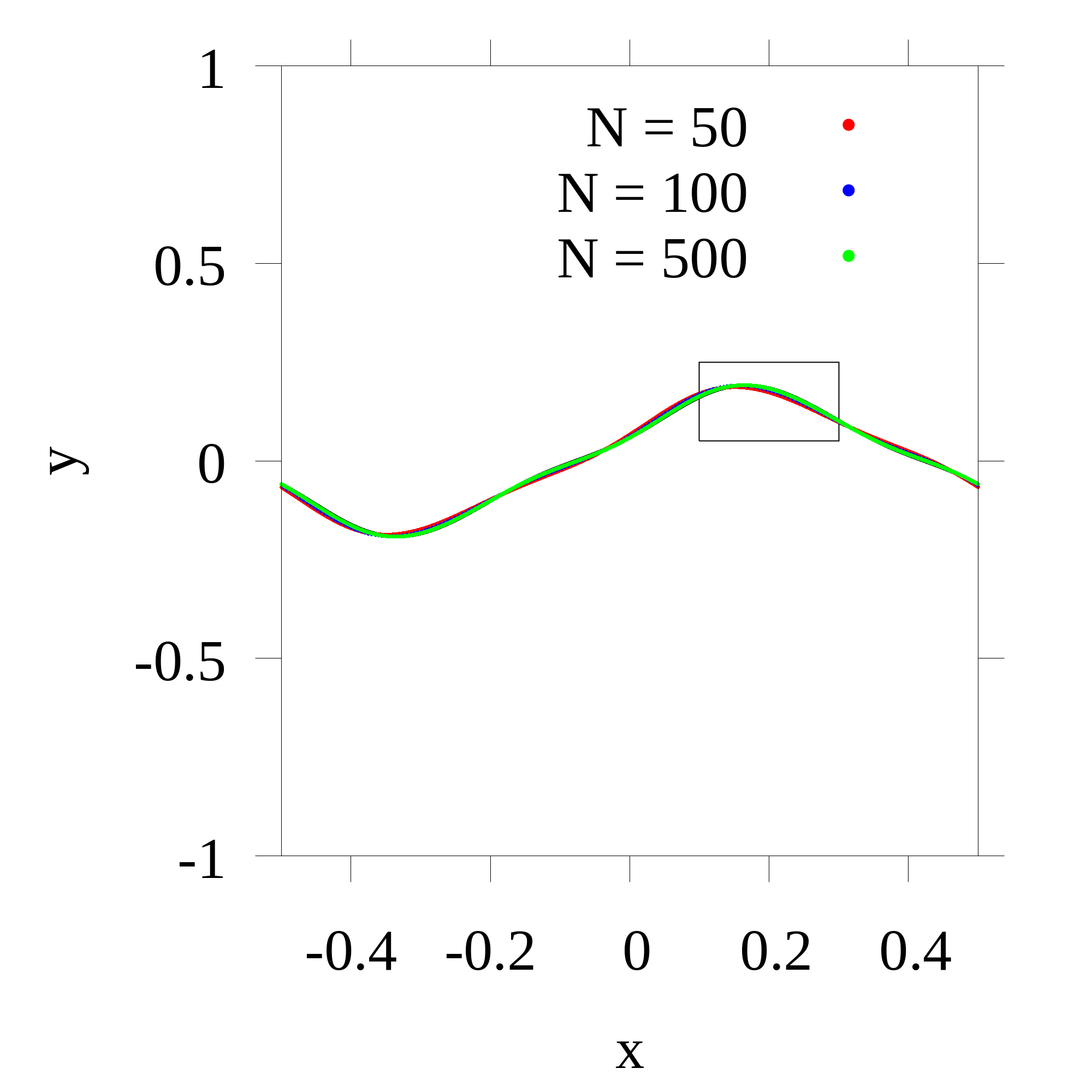}
    \includegraphics[height = 4cm]{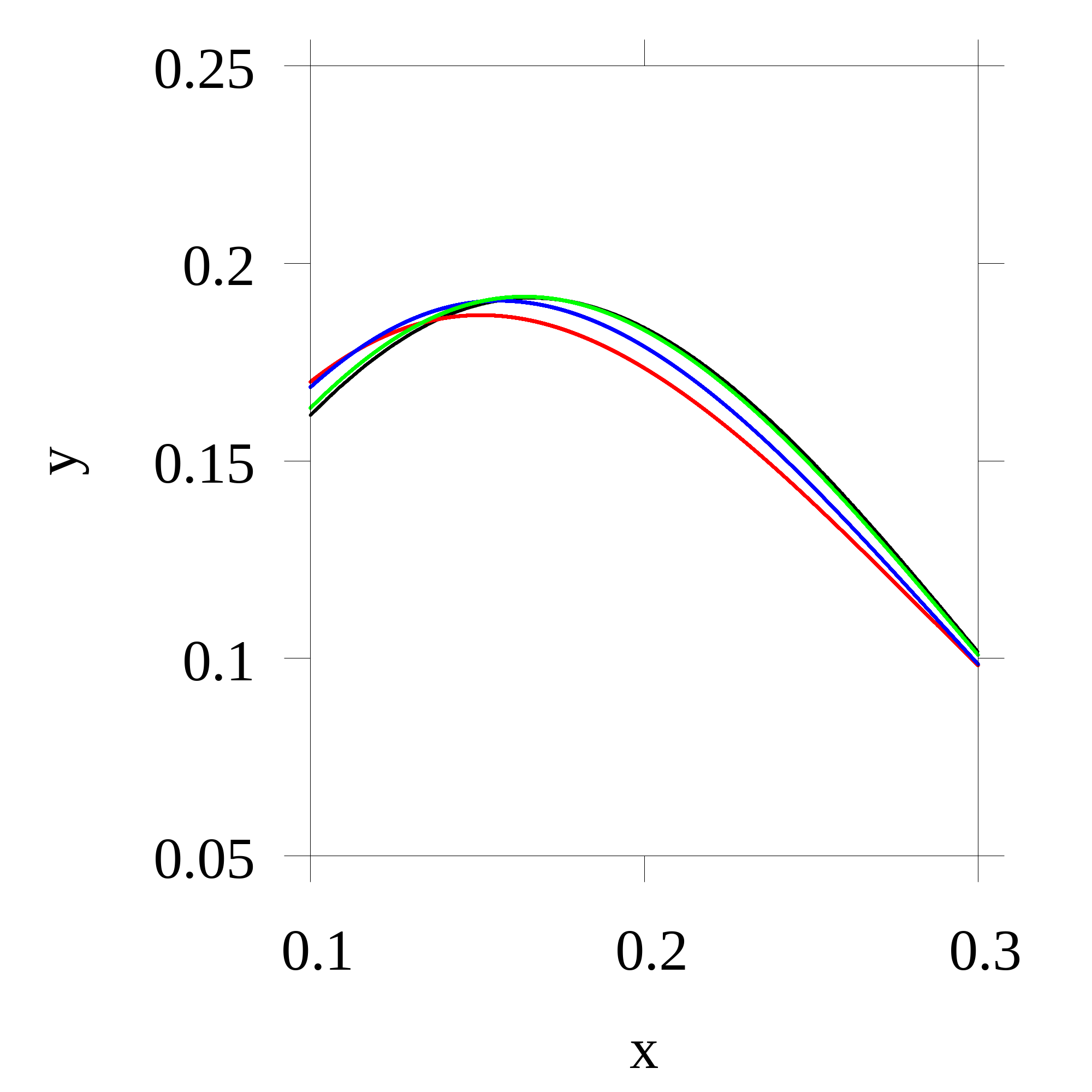}
    \caption{Shearless invariant curve for $b=0.3$ in black and the final shearless snapshot torus for the drift scenario $[b_0=0.0, b_f=0.3]$, $N=50$ in red, $N=100$ in blue and $N=500$ in green, showing that for slower variations of parameter $b_n$ the evolution of NASNM approximates the invariant curves for the static scenario. The right panel shows a close-up of the boxed region in the left panel.}
    \label{adiabatic_portraits}
\end{figure}

Section II was dedicated to showing the main concepts arising from the parameter's time dependence. In this section, we show some results for small and slow parameter variations $b_n$, 

Figure \ref{adiabatic_portraits} shows an evolution scenario for an initial condition with $b_0=0.0$ and $b_f=0.3$. 

We see that the shearless snapshot torus tends to the shearless curve for the static picture for the end of the evolution $b_f = b$. The slower the variation of the parameter $b_n$, the closer the snapshot torus aligns with its corresponding invariant curve in the static picture.

\begin{figure}[H]
    \centering
    \includegraphics[height = 4cm]{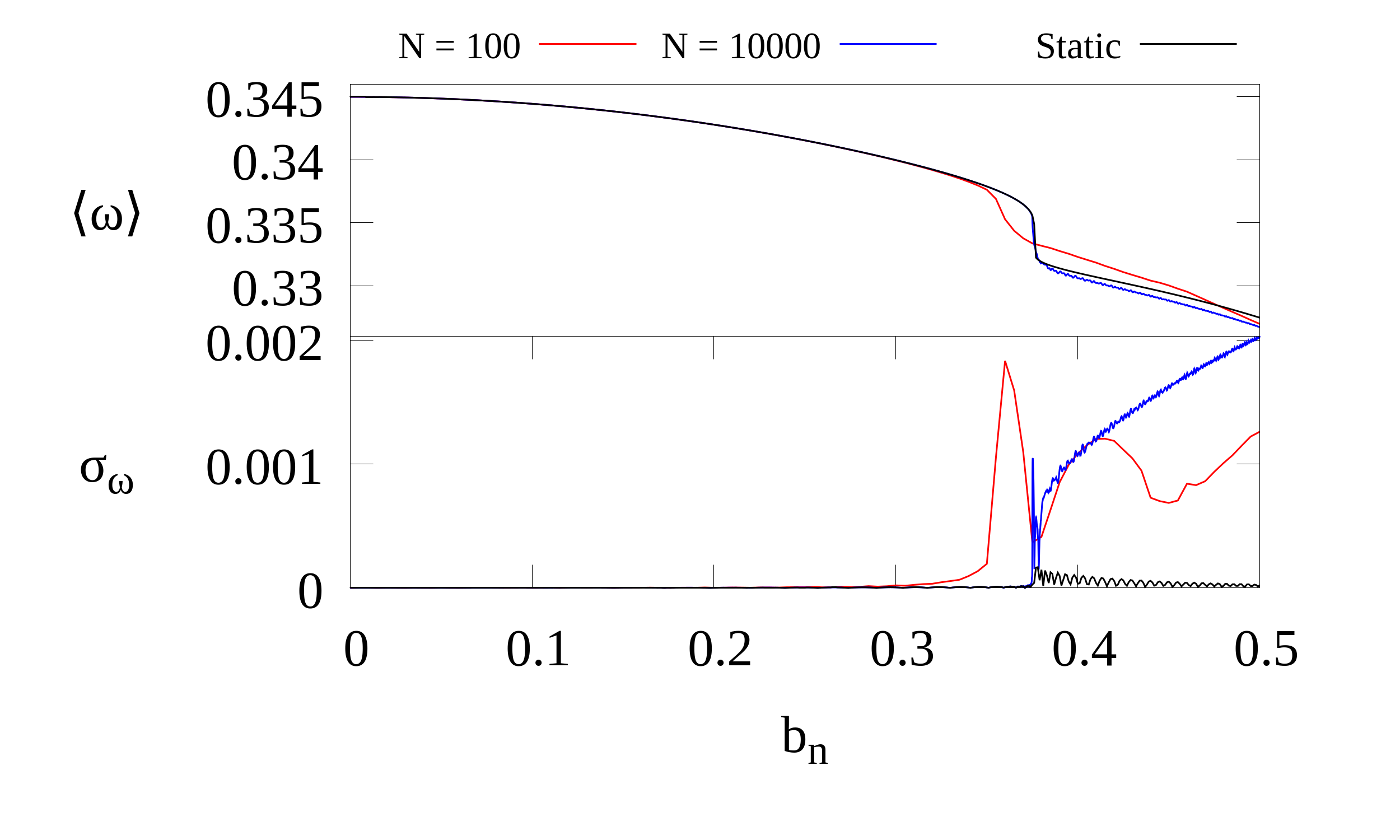} 
    \caption{Averaged rotation number $\langle \omega \rangle$ and the standard deviation $\sigma_\omega$ in function of time-dependent parameter $b_n$. The average is taken over $NP = 10000$ points initially located on the shearless curve for $b_0$. }
    \label{rotation_number_bn}
\end{figure}

We quantify this result by calculating the ensemble-averaged rotation number, namely the average rotation number of all ensemble points initially located on the shearless curve. This result is shown in figure \ref{rotation_number_bn}. 

We see a very good agreement between the ensemble-averaged rotation number and the rotation number in the static scenario for values of $b_n \leq b^*$. But when the parameter reaches values close to the separatrix reconnection $b^*$ (figure \ref{separatrix_reconnection}), the ensemble starts experiencing a stretching towards the separatrix. After that, although the ensemble-averaged rotation number is still close to the rotation number in the static scenario, the phase space is very distinct, notably the standard deviation in figure \ref{rotation_number_bn}. 

\begin{figure}[H]
    \centering
    \includegraphics[height = 4cm]{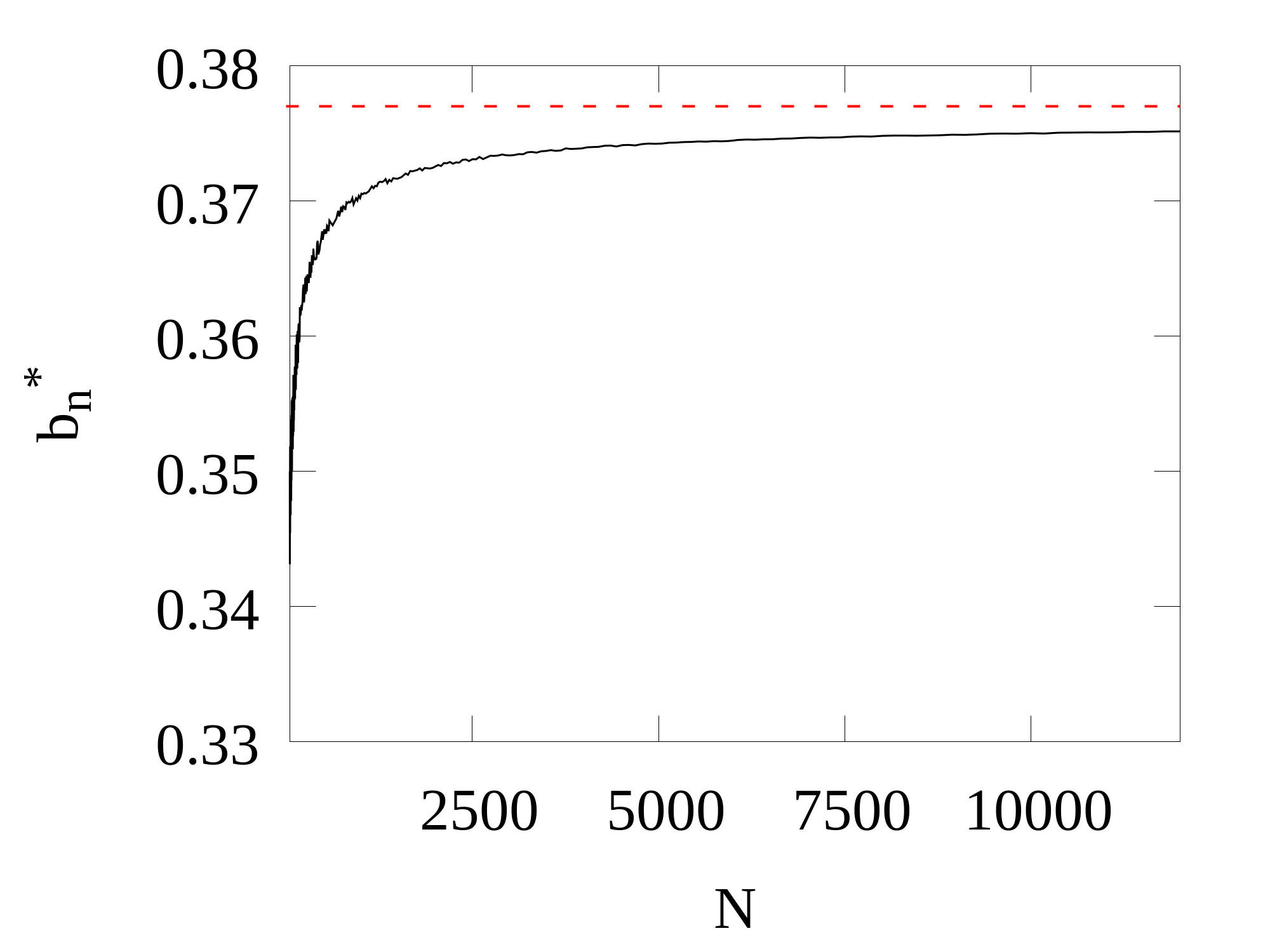} 
    \caption{Critical parameter for separatrix reconnection in function of the total number of iterations. }
    \label{scaling_critical_separatrix}
\end{figure}

For the non-autonomous model, we define the separatrix reconnection threshold $b_n^*$, i.e., the value of $b_n$ where the standard deviation of $\sigma_\omega$ has its maximum value. This result is shown in figure \ref{scaling_critical_separatrix}. 

\section{Shearless snapshot torus breakup}

\begin{figure}[H]
    \centering
    \includegraphics[height = 5cm]{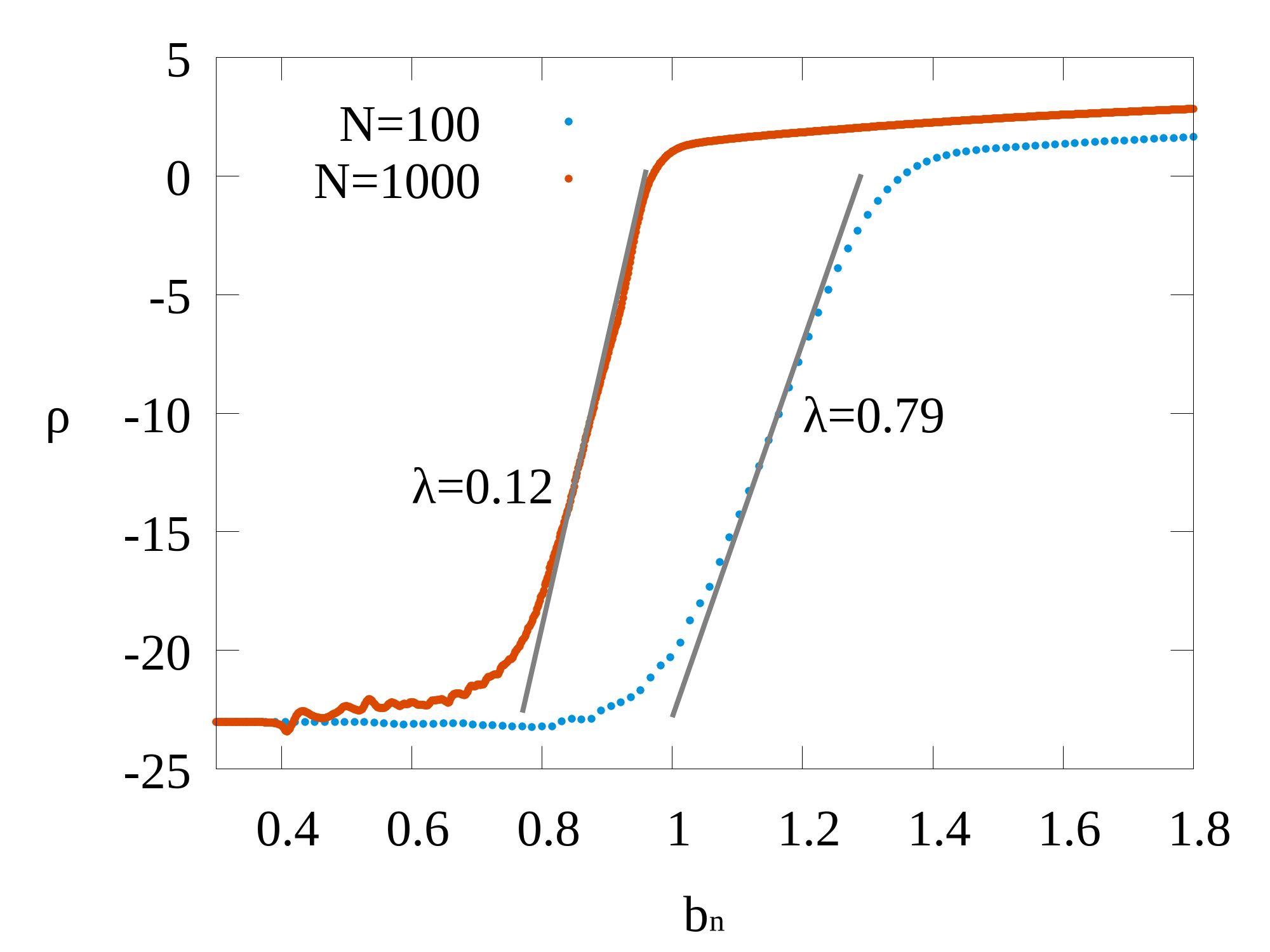} 
    \caption{Ensemble-averaged pairwise logarithmic distance in function of the dynamical parameter $b_n$ for $[b_0=0.3, b_f=1.8]$ and $NP=50000$. In blue $N = 100$, and $N=1000$ in red. }
    \label{eapd}
\end{figure}

This section discusses the transition to chaos of the shearless snapshot torus. For stationary Hamiltonian systems, the transition to chaos is usually associated with the destruction of invariant curves. However, with the introduction of parameter drift, there are no longer invariant curves, so the transition to chaos must be studied with different tools and techniques. To address this, we compute the ensemble-averaged pairwise logarithmic distance (EAPLD) given by the expression \cite{Janosi_D:2024:overview}
\begin{equation}
      \rho(n) = \langle \ln d(n) \rangle,
  \end{equation}
where $d(n)$ is the distance between a pair of points very close to each other in the initial state ($n = 0$). The average $\langle \cdot \rangle$ is calculated over an ensemble of $NP$ pairs of points located initially on the invariant shearless curve for the parameters $(a, b_0)$. Figure \ref{eapd} shows a typical evolution scenario for fixed $[b_0=0.3, b_f=1.8]$ for different numbers of iterations $N$. 

We see a scenario where the distance between the points hardly changes initially. The transition to large-scale chaos begins after a critical parameter $ b_c^-$, where we see an exponentially increasing distance between the points in the ensemble. After the large-scale chaotic transition, for $b_c^+$, the ensemble dynamics become strongly chaotic. The chaotic transition of the shearless snapshot torus is now characterized by the critical parameters $b_c^-$ and $b_c^+$ and the instantaneous Lyapunov exponent $\lambda$, which is defined as the slope of the curve during the exponential transition, calculated in function of the discrete-time $n$. The calculated instantaneous Lyapunov exponent $\lambda$ is unique because it exclusively characterizes the shearless snapshot torus. This parameter can be ascertained only after a certain duration following the beginning of the chaotic transition.

We observe that the chaotic transition starts at lower values of the parameter $b_n$ when the parameter changes more slowly. This is attributed to the fact that a slower change brings us closer to the stationary scenario, where chaos in a region of the phase space is closely tied to the presence of invariant curves. As the growth rate of the parameter increases, we move further away from the stationary scenario, and it takes some time for our system to begin experiencing significant stretching, ultimately leading to chaotic behavior.

Although the phase space in non-autonomous systems depends on the evolution scenario of the parameter $[b_0, b_f, N]$, the critical parameters for the chaotic transition $b_c^-$ and $b_c^+$ are associated with the stationary behavior of the system. In a stationary state, only one critical parameter $b_c$ defines the shearless breaking point. The invariant curve is broken once this point is reached, but partial barriers may still exist (see refs \cite{Szezech_JD:2009:Transport_properties_in_nontwist_area-preserving_maps, Szezech_JD:2012:Effective_transport_barriers_in_nontwist_systems}). In a time-dependent system, the transition is smooth and influenced by $N$. 

\begin{figure}[H]
    \centering
    \includegraphics[height = 5cm]{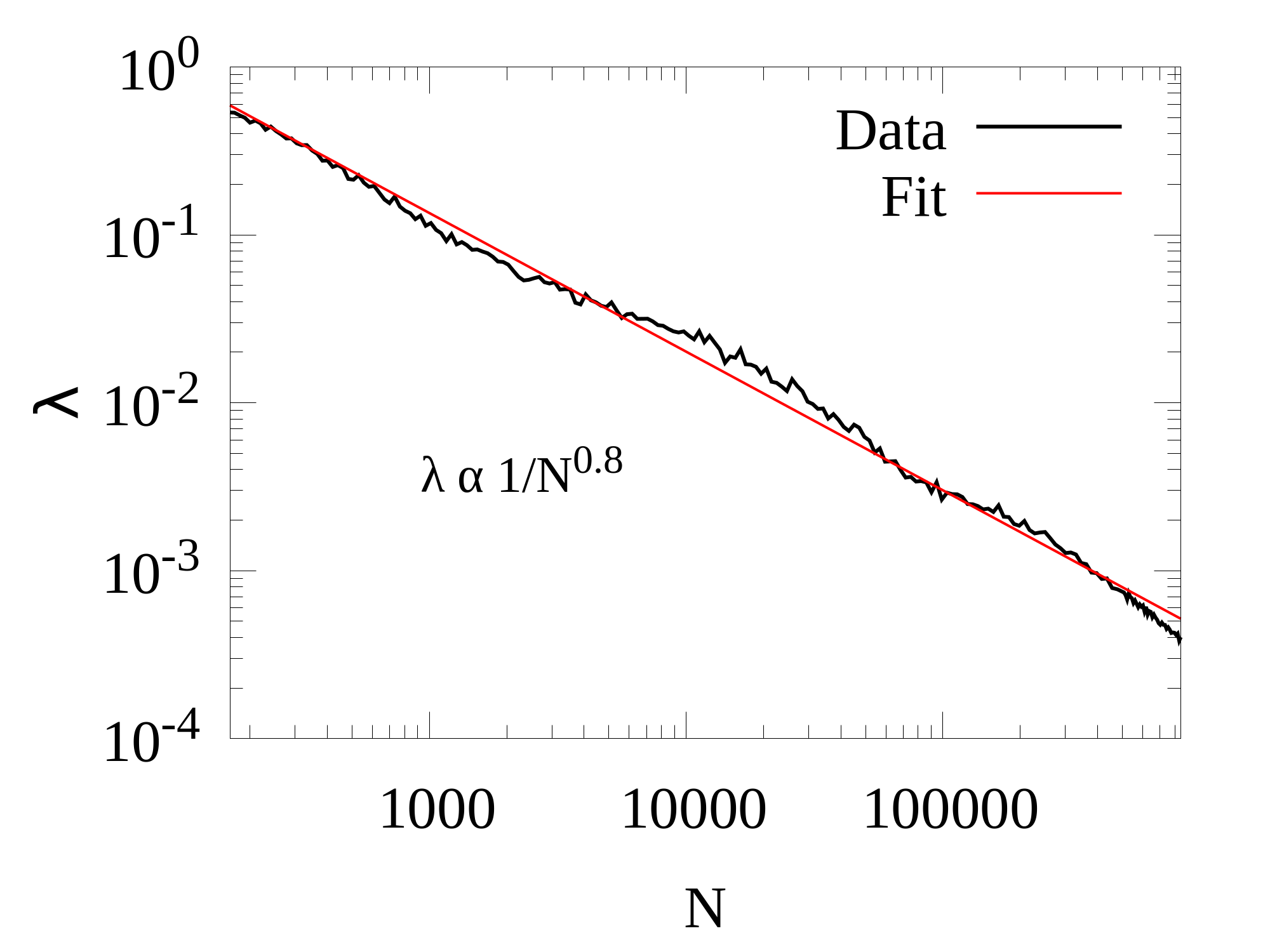}
    \includegraphics[height = 5cm]{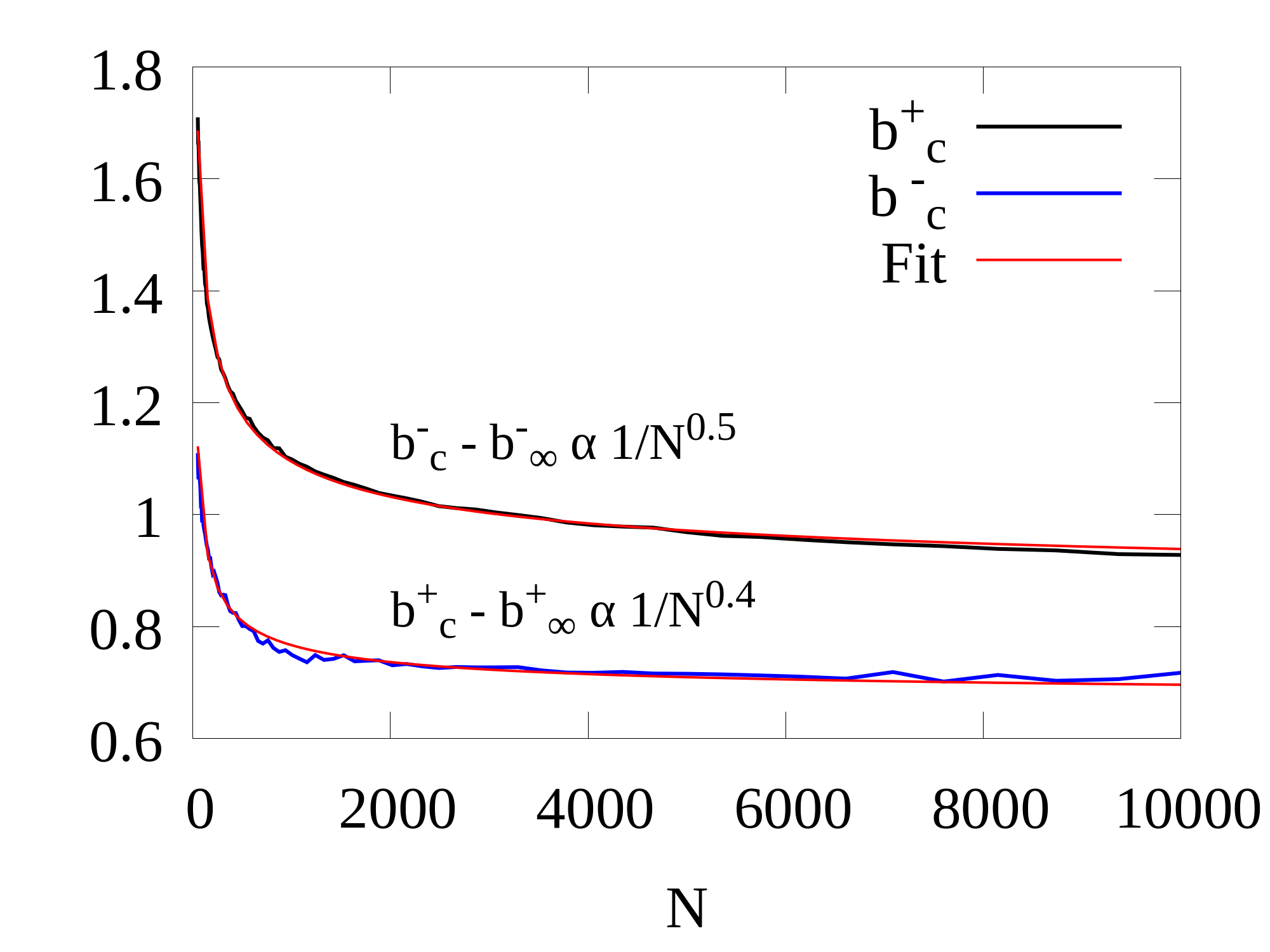}
    \caption{Scaling for the instantaneous Lyapunov exponent $\lambda$ and the critical parameter $b_c$ in function of number of iterations $N$. }
    \label{eapd_scalings}
\end{figure}

Figure \ref{eapd_scalings} shows the scalings between $\lambda$ and $b_c$ with the total number of iterations for $b_0=0.3$ and $b_f=1.8$. 

We see that for slower variations of the parameter $b_n$, the critical parameters $b_c$ have an asymptotic behavior. The Lyapunov exponent tends to zero as the number of iterations tends to infinity because, for conservative Hamiltonian systems, a volume element of the phase space will stay the same along a trajectory and remnants of KAM tori hinder transport \cite{Meiss92} . 

\begin{figure}[H]
    \centering 
    \includegraphics[height = 6cm]{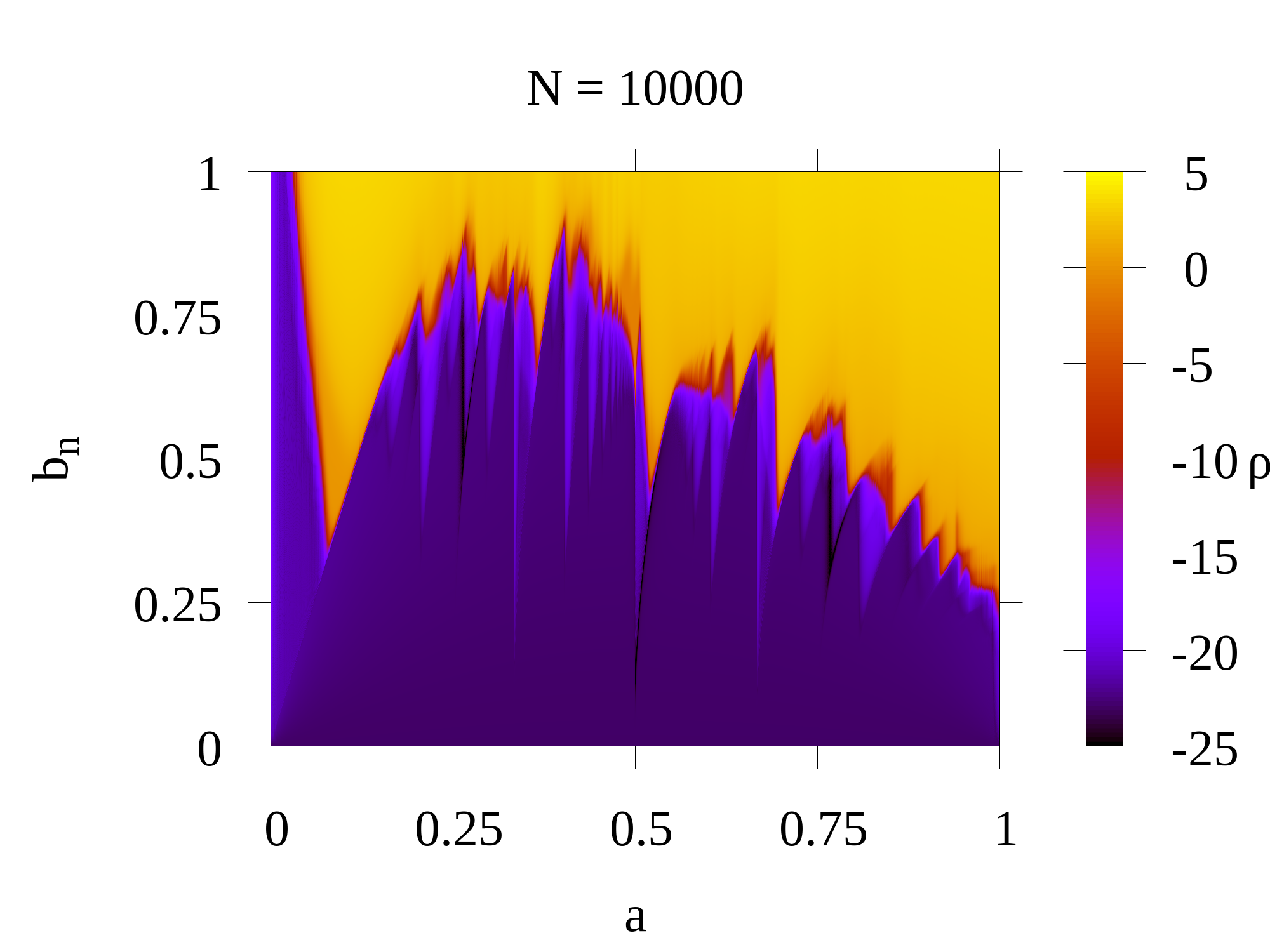} 
    \caption{Parameter space for the non-autonomous SNM. The color scheme represents the ensemble-averaged pairwise logarithmic distance. }
    \label{eapd_PS}
\end{figure}

Figure \ref{eapd_PS} shows the ensemble-averaged pairwise logarithmic distance calculated on the parameter space. We do the following: for each (fixed) value of parameter $a$, we define the initial condition as points on the shearless curve for $b_0=0$, and we iterate it using the NASNM, for each instant, we calculate the EAPLD, showed through the color. Although the phase spaces are very different from the static case to the non-autonomous, we see the similarities in the parameter space with the static case (see ref. \cite{wurm2004reconnection}). 

\begin{figure}[H]
    \centering
    \includegraphics[height = 6cm]{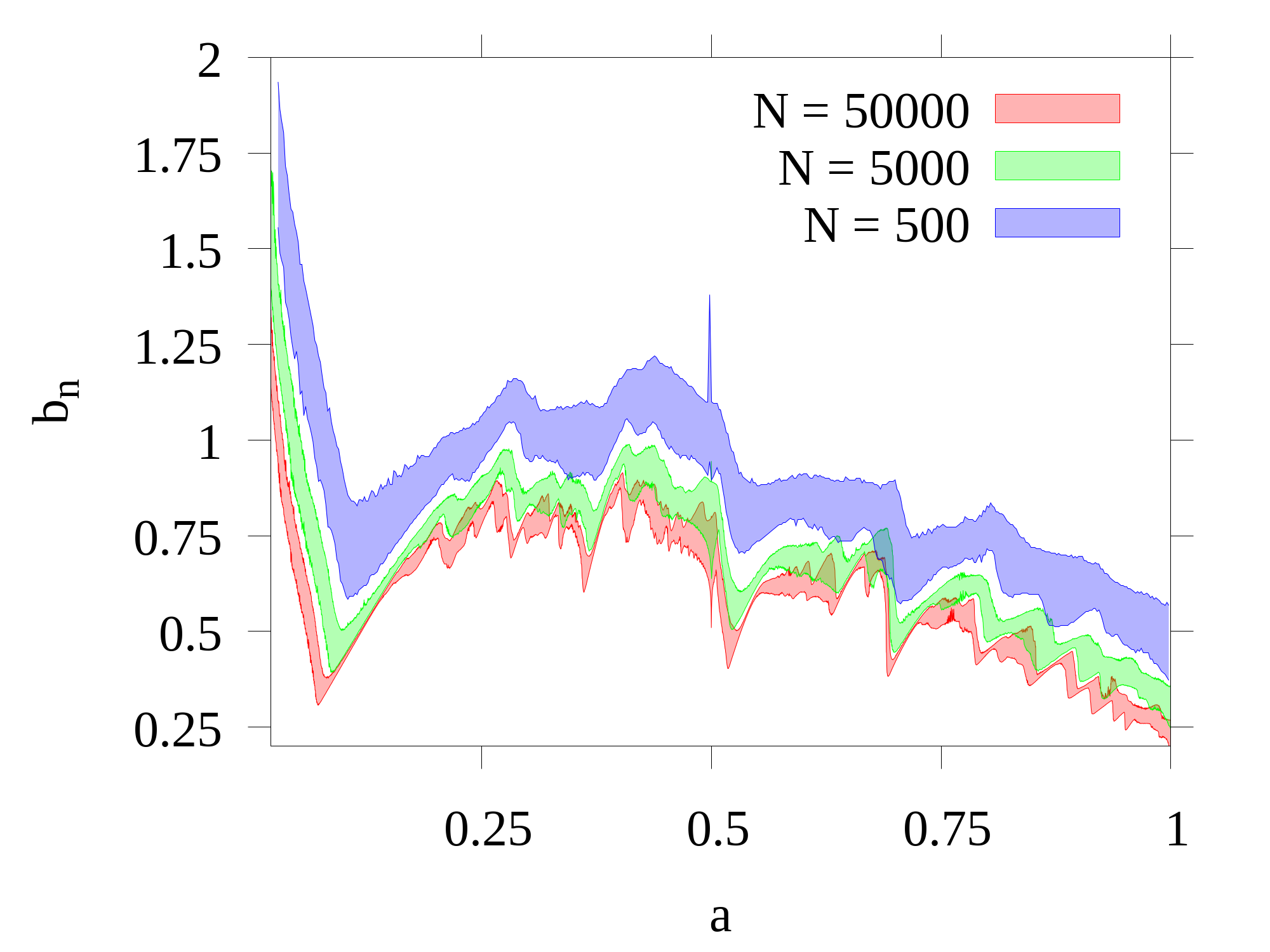}   
    \caption{Transient regions visualized on the parameter space for different numbers of iterations.}
    \label{eapd_PS_transition}
\end{figure}

Figure \ref{eapd_PS_transition} shows the transient region $b_c^- < b_n < b_c^+$ for different numbers of iterations. We see that slower parameter growth rate leads to a sharper transition, closer to the static picture, while faster parameter growth rate exhibits a smoother transition. 

As we discussed in section 1, the SNM is reversible, implying that we can represent the time-reversed map $M^{(-n)}$ as

\begin{equation}
    \begin{split}
      &     y_{n} = y_{n + 1} + b_n \sin(2\pi x_n), \\ 
      &     x_{n} = x_{n + 1} - a (1 - y_{n + 1}^2) 
    \end{split}
    \label{NASNM-reversed}
\end{equation}

Now let us define the initial state  ${\mathbf{R}}_0$ as a sequence of points initially on the shearless curve for $b_0 = 0$. The ensemble evolution is followed through the evolution scenario $[b_0=0, b_f=1, N]$, resulting in the final state ${\mathbf{R}}_N = M^{N} ({\mathbf{R}}_0) $. Then, we reconstruct the initial ensemble by applying the time-reversed NASNM \eqref{NASNM-reversed}, denoted by $\mathbf{R}_{0'} = M^{(-N)} (\mathbf{R}_N)$.

The reversibility error $D$ is measured as the average distance between the points of the initial ensemble ${\mathbf{R}}_0$ and the reconstructed initial ensemble $\mathbf{R}_{0'}$: 

\begin{equation}
    D(\mathbf{R}_0, \mathbf{R}_{0'}) = \langle |\mathbf{r}_{0'} - \mathbf{r}_0| \rangle 
\end{equation}

\begin{figure*}
    \centering
     \includegraphics[height = 5.5cm]{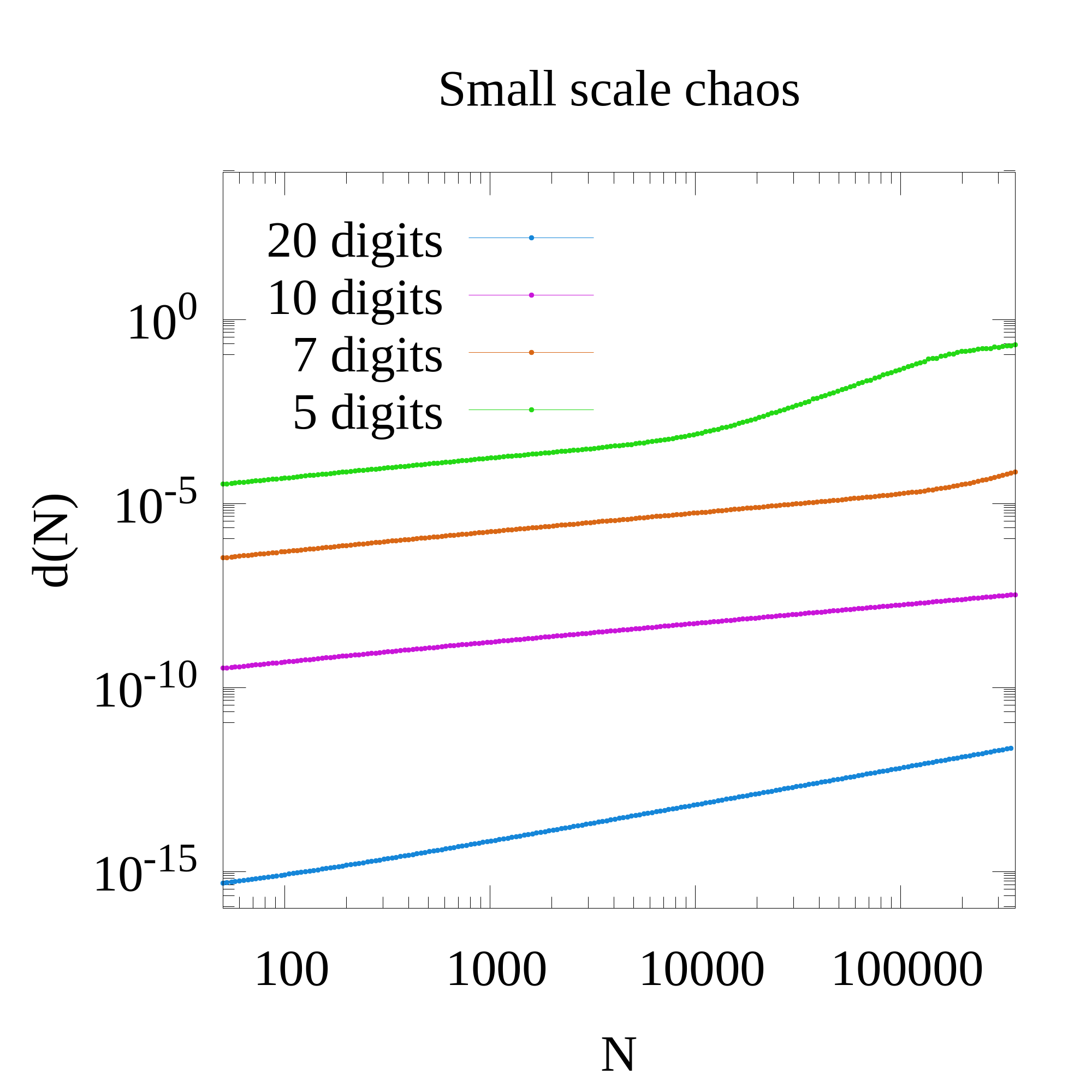}
     \includegraphics[height = 5.5cm]{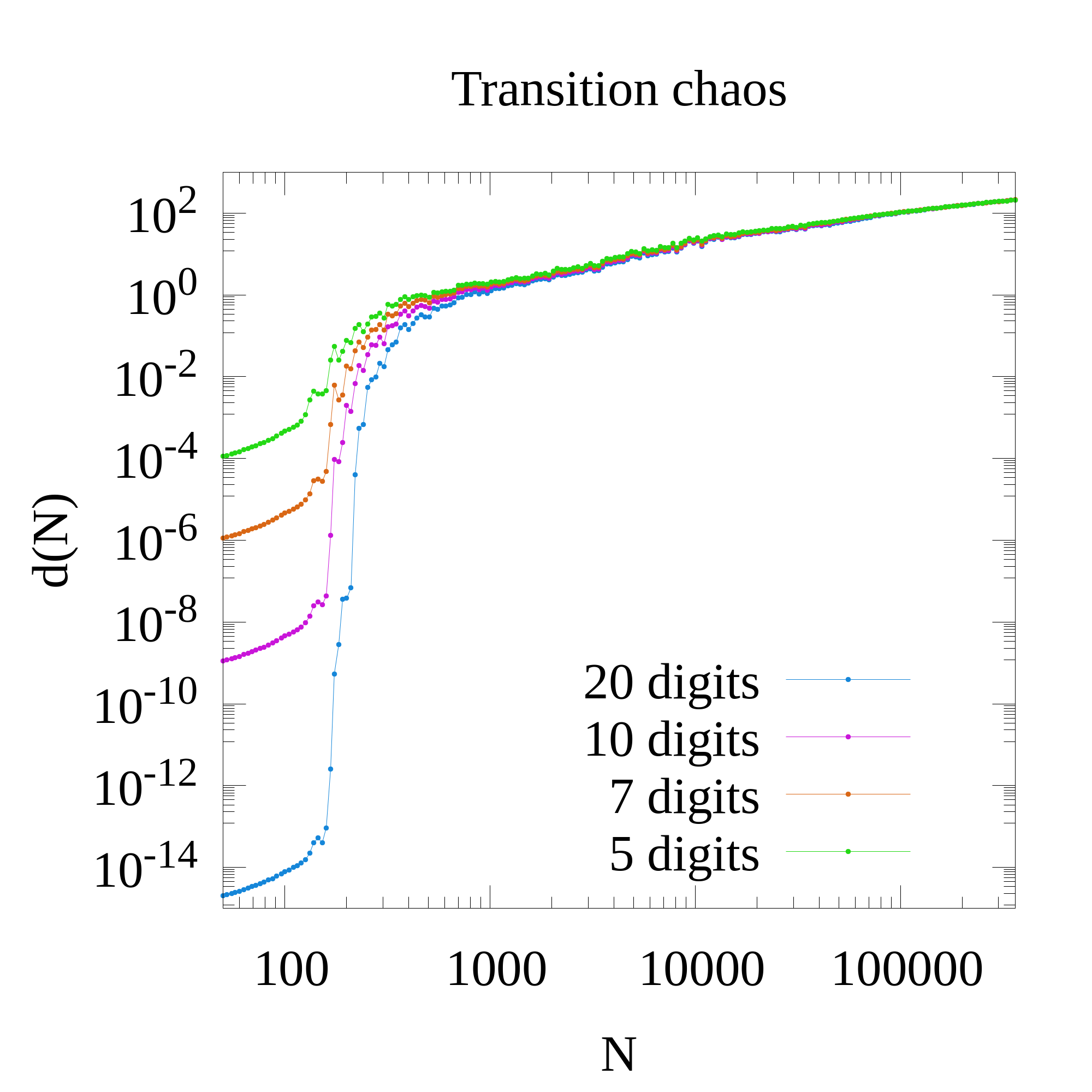}
     \includegraphics[height = 5.5cm]{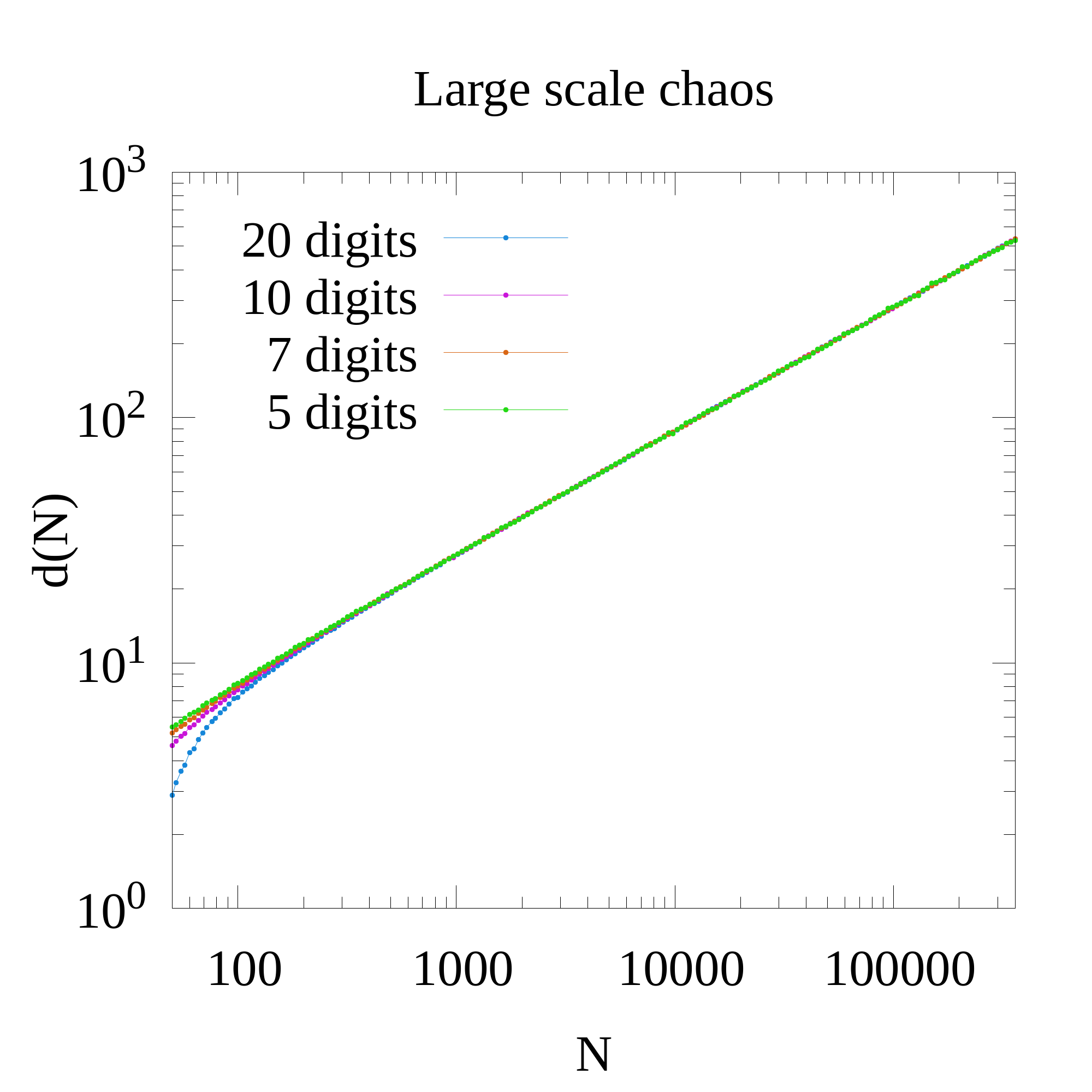}
    \caption{Reversibility error in function of $N$ in three different scenarios: small-scale chaos corresponds to $[b_0=0, b_f=0.3]$, the intermediate range is $[b_0=0.3, b_f=0.8]$ and finally, large-scale chaos is $[b_0=0.3, b_f=1.8]$. The different colors represent the number of precision digits used for the calculations. }
    \label{reversibility.png}
\end{figure*}

The reversibility error calculated under different scenarios of parameter evolution provides an additional tool for examining the transition to large-scale chaos in our system. Figure \ref{reversibility.png} shows the reversibility error in function of $N$ in three different ranges of the parameter $[b_0, b_f]$: in (a) we consider the case where small scale chaos is present, in (b) the parameter range is such that the chaotic transition might occur, depending on the parameter`s growth rate and (c) shows the result for a parameter range such that large-scale chaos is found. 

\section{Transport properties}

In this section, we briefly discuss the transport properties of the NASNM. Unlike the SNM with constant parameters, where we see different scenarios of collision/annihilation of periodic orbits leading to global transport, depending on the region in the parameter space, for the SNM with time-dependent parameters, the route to chaos is not only related to the two parameters $(a,b)$ but also to the scenario of the evolution of parameter $b_n$. 

\begin{figure*}
    \centering
    \includegraphics[height = 4cm]{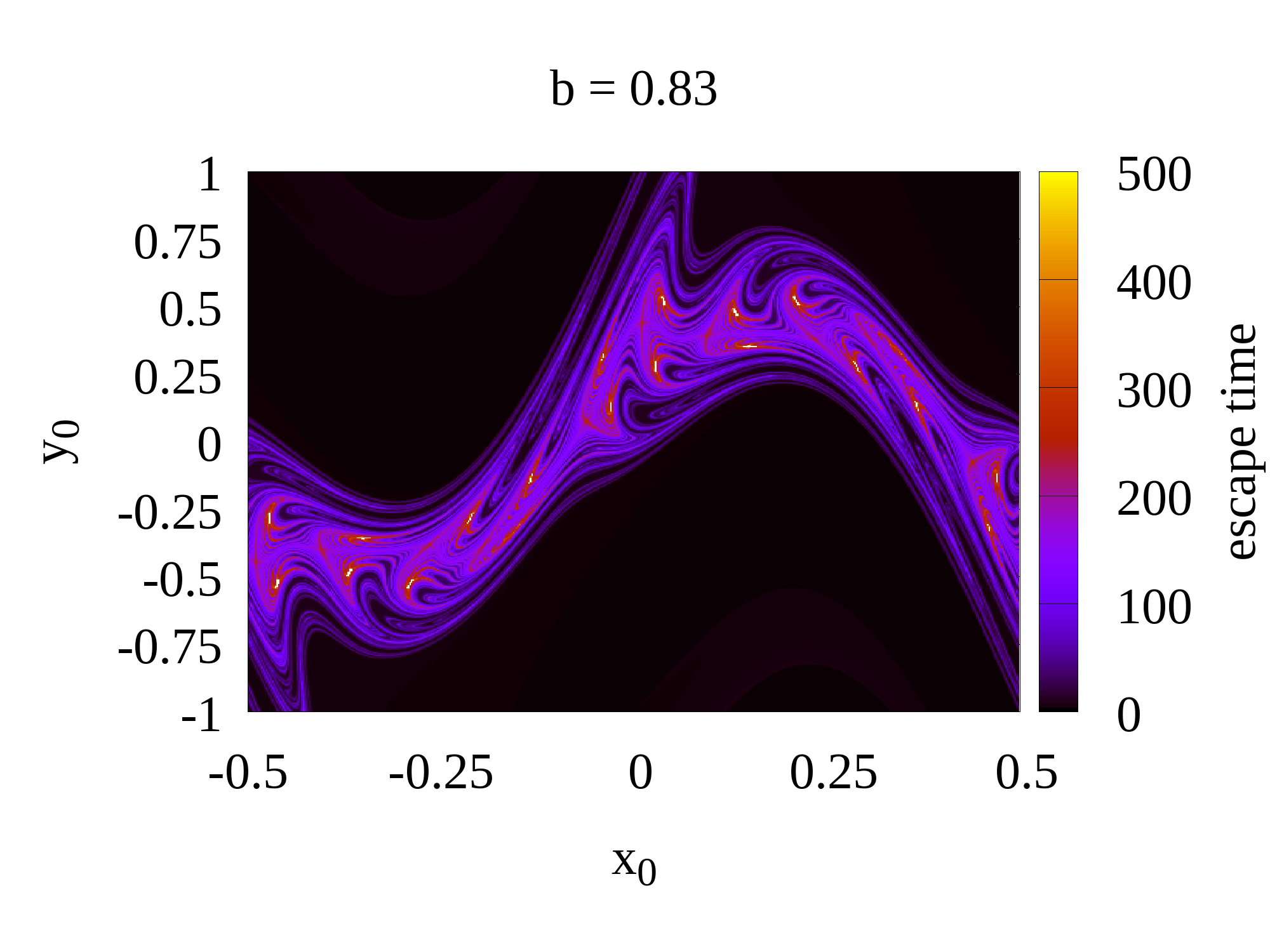}
    \includegraphics[height = 4cm]{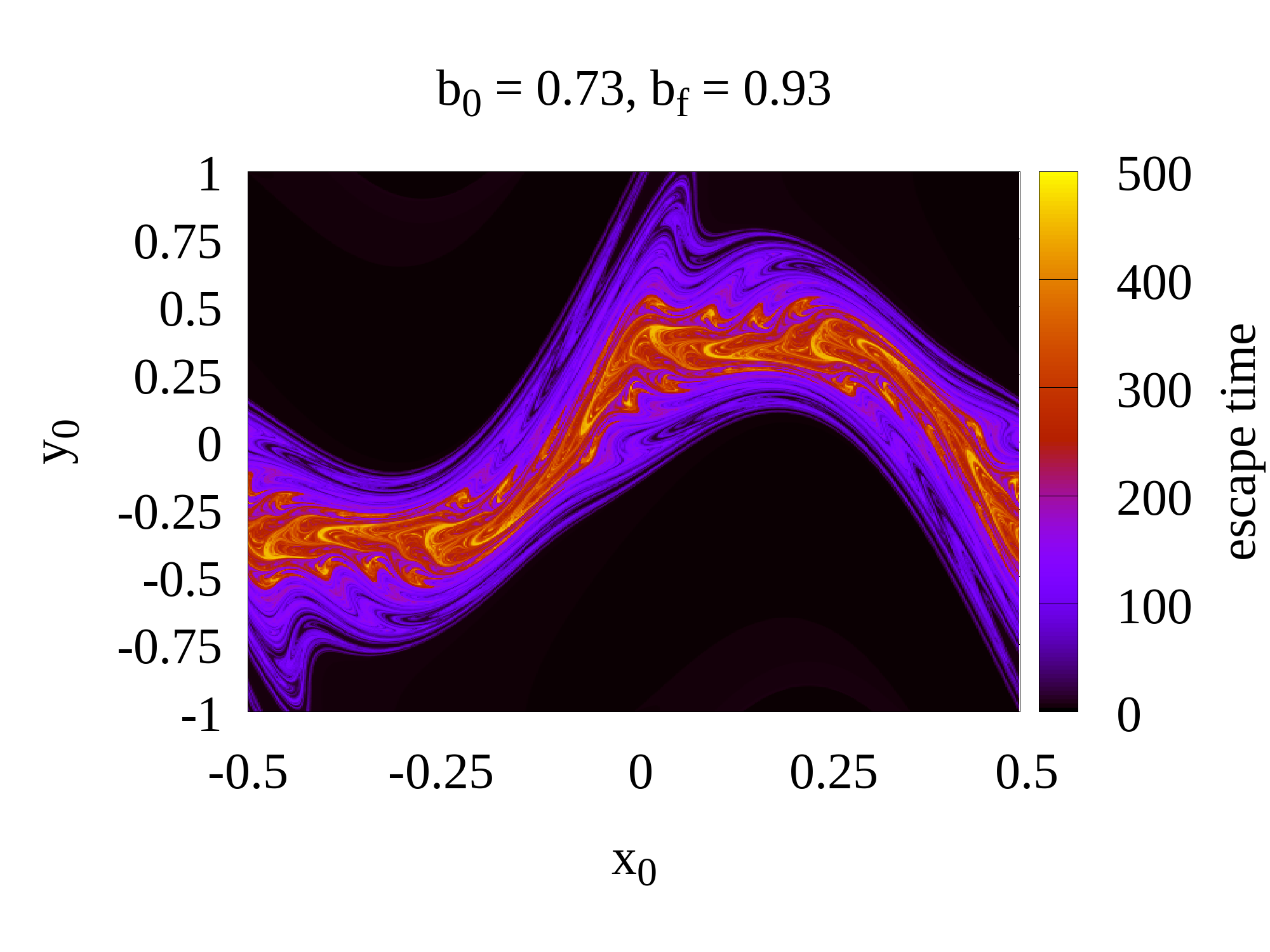}
    \includegraphics[height = 4cm]{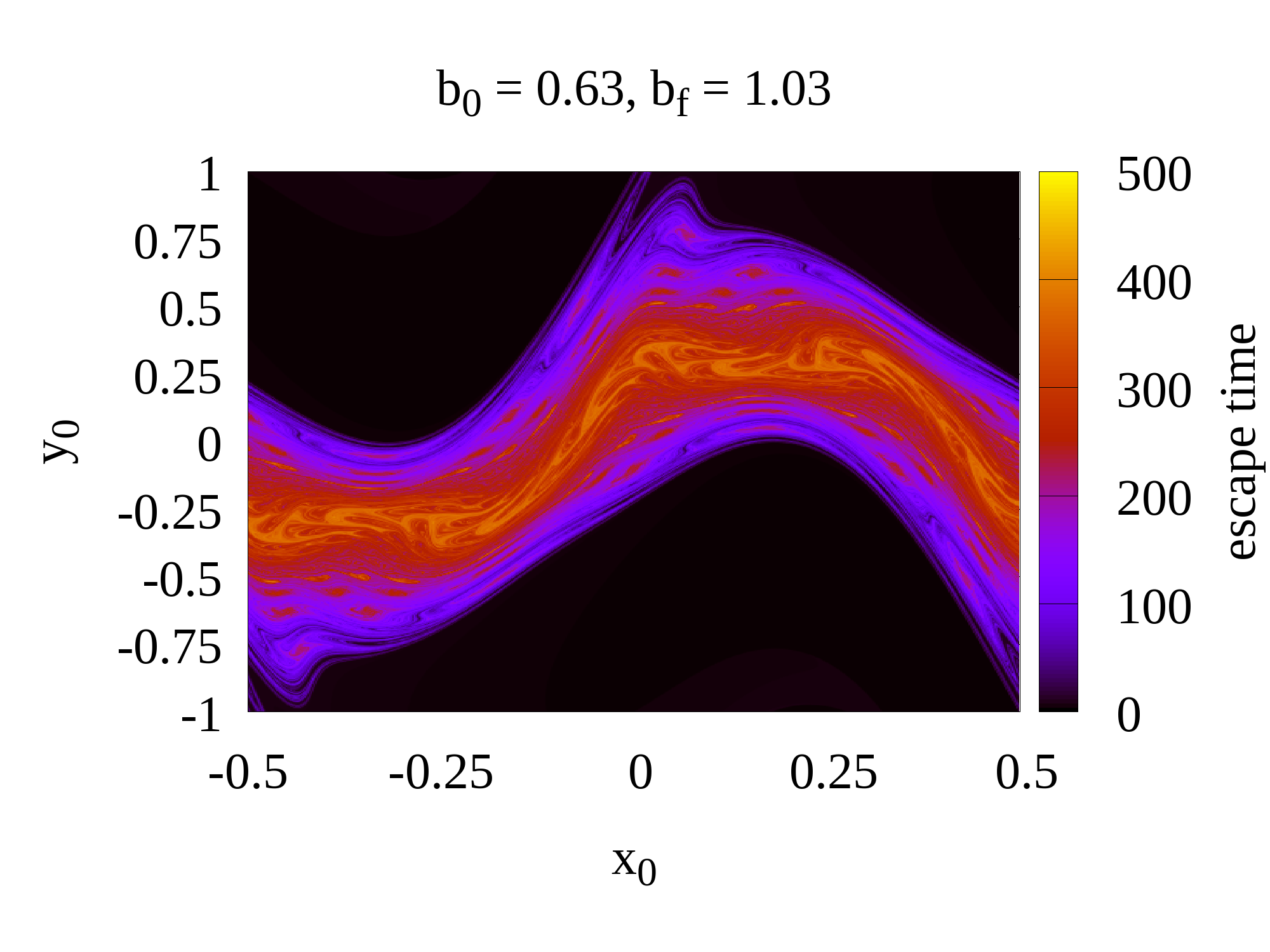}
    \caption{Escape time on the initial phase space for the stationary case, on the left, and for the non-autonomous scenario in the center and on the right, for $\Delta b = 0.1$ and $\Delta b = 0.2$, respectively. We considered the threshold value $y_{TS} = 1$. }
    \label{Fn}
\end{figure*}

We start this section by defining the escape time. The escape time of an orbit $(x_n, y_n)$ is defined as the number of iterations needed to reach a condition of escape $y_n \geq y_{TS}$ in the radial coordinate. Figure \ref{Fn} shows the escape time on the phase space for the stationary map and the non-autonomous map with different evolution scenarios of the parameter. The stationary SNM, with the critical parameter $b=b_c=0.83$, and the NASNM, with fixed total number of iterations $N$ and different scenarios for $[b_0 = b_c - \Delta b,b_f = b_c + \Delta b]$, where $b_c$ is the critical parameter for the shearless breaking point in the stationary model.   

In the stationary picture, for $b_c = 0.83$, the shearless barrier is already destroyed, but we can still see a partial barrier where the shearless curve was previously. We see a region where orbits stay longer, an effect called stickiness. 

The non-autonomous map shows a scenario where the region around the shearless barrier also acts as a barrier, as we see by the longer escape time within this area. That is because the system preserves a memory of the parameter's evolution. For instance, for $b_0 = 0.73$, the central barrier still exists on the stationary picture; therefore, orbits in the central region start escaping after the time they would if we start with $b_0=0.83$, where the shearless barrier does not exist anymore on the stationary map. 

Let us now consider the average square radial displacement, $\langle \Delta y(n) ^ 2 \rangle$ 

\begin{equation}
    \langle \Delta y(n) ^ 2 \rangle = \langle (y(n) - y(0) ) ^2 \rangle = \frac{1}{NP} \sum_{i=1}^{NP} (y_i(n) - y_i(0) ) ^ 2,  
\end{equation}
where the average is taken over an ensemble of $NP$ points, each $i$-th point located initially at $(x_i(0), y_i(0))$. 

\begin{figure}[H]
    \centering
    \includegraphics[height = 5cm]{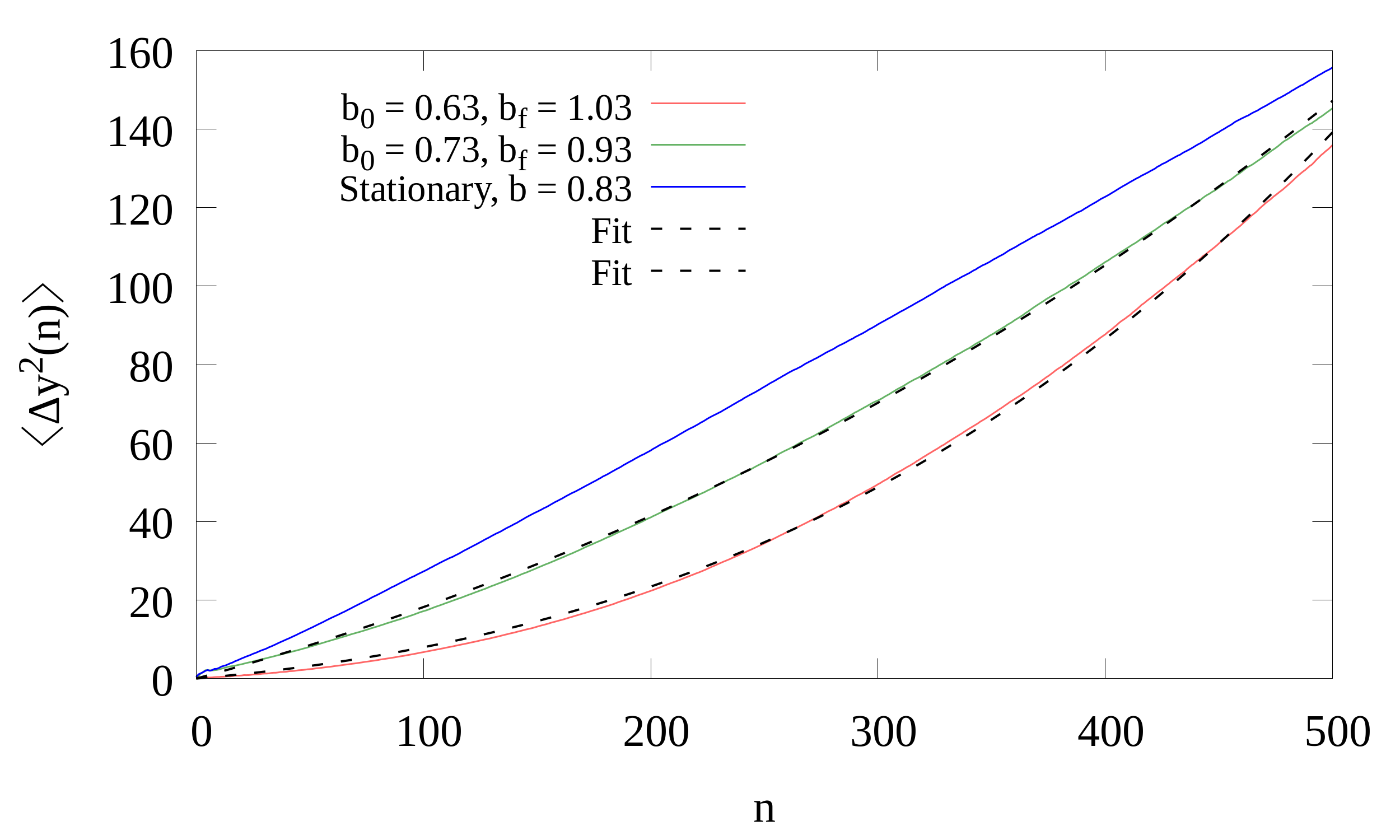}
    \caption{Average square radial displacement in function of discrete-time $n$, for an initial ensemble of $NP = 10^5$ points located at the horizontal line $y = 0$. The curve in blue represents the static map with $b=b_c=0.83$. The curves in red and green represent the non-autonomous map for $\Delta b = 0.2$ and $\Delta b = 0.1$, respectively. The dotted curves represent fits using the random walk approach with time-increasing steps. }
    \label{msd}
\end{figure}

Figure \ref{msd} shows the average square radial displacement in function of  $n$ for the stationary SNM, with the critical parameter $b=b_c=0.83$, and the NASNM, with fixed total number of iterations $N$ and different scenarios for $[b_0 = b_c - \Delta b,b_f = b_c + \Delta b]$.  
 
For the static map with $b \geq b_c$, curve in blue in figure \ref{msd}, the shearless curve is broken, large-scale chaos is dominant and the dynamics on phase space is approximated as a two-dimensional random walk with independent steps, where $x_n$ is a sequence of independent random variables distributed uniformly on $[-1/2, 1/2]$ (and the increments $(y_{n+1} - y_n)$ are accordingly independent, with the resulting arcsine distribution). We see that the average square radial displacement for the stationary map converges towards a Gaussian diffusion \cite{Chirikov:1979}, with the equation $\langle \Delta y(n)^2 \rangle = (b^2/2) n$. 

For the non-autonomous map, curves in red and green, diffusion behaves differently. Initially, when \( b_n \leq b_c \), diffusion is weaker, but as the parameter increases for larger values of \( n \), the diffusion is progressively enhanced, with \(\langle \Delta y(n)^2 \rangle\) increasing accordingly. Using a random walk approach with time-increasing step sizes, we have fitted the calculated values of \(\langle \Delta y(n)^2 \rangle\), confirming this enhancement. This effect becomes more pronounced as the system diverges from the static scenario, and by fixing \( N \) while taking \( \Delta b \to 0 \), the static behavior is recovered.

\section{Conclusions}

This paper introduced the non-autonomous standard nontwist map by considering a drift on parameter $b$. Conserved quantities and invariant structures, such as KAM tori, are lost for maps with time-dependent parameters. Because of that, we study the system's dynamics by following a set of trajectories that initially start in a single torus of the initial system; in this paper, we followed trajectories initially along the shearless curve. Although there are no invariant structures, we show that the ensemble evolution follows its respective torus in the static map, i.e., the ensemble points are located in the same line in the phase space and have the same rotation number. But after the parameter $b_n$ crosses the separatrix reconnection threshold, the ensemble starts experiencing a strong stretching because the ensemble points are located in a chaotic domain associated with a heteroclinic tangle of the autonomous dynamics. 

In the context of non-autonomous systems, the transition to chaos of the shearless barrier was shown to be such that the points of the set initially hardly deviate from each other. The transition begins after a critical value of the parameter \( b^-_c \). During the transition, the points of the set diverge exponentially from each other, with a growth rate of the distances characterized by the instantaneous Lyapunov exponent \( \lambda \). For \( b^+_c \), the transition ceases, the distance between the points saturates, and the remnants of the shearless set exhibits chaotic dynamics. Finally, we present power laws relating the parameters \( \lambda \), \( b^-_c \), and \( b^+_c \) as functions of the total iteration time \( N \). We show the asymptotic behavior of the critical parameters, whereas the instantaneous Lyapunov exponent tends to zero. A second analysis related to the chaotic transition was provided, where we study the distance between the initial condition and the reconstructed initial condition, found by iterating backwards the final state with the time-reversed map. We show that the initial condition can be well reconstructed for a parameter variation scenario where small-scale chaos is found. 

Finally, we demonstrate that the parameter drift introduces an additional transport mechanism, altering the diffusion dynamics of the system. In the static case, where \( b \geq b_c \), the diffusion is Gaussian and consistent with a random walk characterized by independent steps. For the non-autonomous case, diffusion initially occurs more slowly when the parameter is below the critical threshold. However, as the parameter increases over time, the diffusion is enhanced, leading to greater radial displacements. In analogy to the stationary case, we show that the diffusion properties of the non-autonomous map, for values of $b_n$ large enough, is approximated by the random walk with time-increasing step sizes. 

 
\section*{Acknowledgments}

The authors thank the financial support from the Brazilian Federal Agencies (CNPq) under Grant Nos. 407299/2018-1, 302665/2017-0, 403120/2021-7, and 301019/2019-3, and the São Paulo Research Foundation (FAPESP, Brazil) under Grant Nos. 2018/03211-6, 2022/04251-7 and 2022/05667-2; and support from Coordenação de Aperfeiçoamento de Pessoal de Nível Superior (CAPES) under Grants No. 88887.710886/2022-00, 88887.522886/2020-00,88881.143103/2017-01 and Comité Français d’Evaluation de la Coopération Universitaire et Scientifique avec le Brésil (COFECUB), Brazil under Grant No. 40273QA-Ph908/18. Centre de Calcul Intensif d’Aix-Marseille is acknowledged for granting access to its high-performance computing resources. YE enjoyed the hospitality of the grupo controle de oscilações at USP.


\begin{thebibliography}{10}
\expandafter\ifx\csname url\endcsname\relax
  \def\url#1{\texttt{#1}}\fi
\expandafter\ifx\csname urlprefix\endcsname\relax\def\urlprefix{URL }\fi
\expandafter\ifx\csname href\endcsname\relax
  \def\href#1#2{#2} \def\path#1{#1}\fi

\bibitem{Lichtenberg_AJ:2ed:Regular_and_chaotic_dynamics}
A.~J. Lichtenberg, M.~A. Lieberman, Regular and chaotic dynamics, Springer, New York, 1992.

\bibitem{Lowenstein_JH:2012:Essentials_of_Hamiltonian_Dynamics}
J.~H. Lowenstein, Essentials of Hamiltonian Dynamics, Cambridge University Press, Cambridge, 2012.

\bibitem{Portela_JSE:2007:Diffusive_transport_throught_a_nontwist_barrier_in_tokamaks}
J.~S. Portela, I.~L. Caldas, R.~L. Viana, P.~Morrison, Diffusive transport through a nontwist barrier in tokamaks, International Journal of Bifurcation and Chaos 17~(05) (2007) 1589--1598.

\bibitem{Hazeltine_JD:2003:Plasma_Confinement}
R.~D. Hazeltine, J.~D. Meiss, Plasma Confinement, Dover Publications, New York, 2003.

\bibitem{Ottino_JM:1989:The_Kinematics_of_Mixing}
J.~M. Ottino, The Kinematics of Mixing: Stretching, Chaos, and Transport, Cambridge Texts in Applied Mathematics, Cambridge University Press, Cambridge, 1989.

\bibitem{Pierrehumbert_RT:1991:Large-scale_horizontal_mixing_in_planetary_atmospheres}
R.~T. Pierrehumbert, {Large‐scale horizontal mixing in planetary atmospheres}, Physics of Fluids A: Fluid Dynamics 3~(5) (1991) 1250--1260.

\bibitem{Kyner_WT:1973:Invariant_manifolds_in_celestial_mechanics}
W.~T. Kyner, Invariant manifolds in celestial mechanics, in: B.~D. Tapley, V.~Szebehely (Eds.), Recent Advances in Dynamical Astronomy, Springer Netherlands, Dordrecht, 1973, pp. 192--196.

\bibitem{Sousa_MC:Energy_exchange_coupled_system}
M.~{C. de Sousa}, A.~Schelin, F.~Marcus, R.~{L. Viana}, I.~{L. Caldas}, Internal energy exchanges and chaotic dynamics in an intrinsically coupled system, Physics Letters A 453 (2022) 128481.

\bibitem{Jánosi_D:2019:Chaos_In_hamiltonian_systems_subjected_to_parameter_drift}
D.~J{\'a}nosi, T.~T{\'e}l, Chaos in hamiltonian systems subjected to parameter drift, Chaos: An Interdisciplinary Journal of Nonlinear Science 29~(12) (2019) 121105.

\bibitem{Jánosi_D:2021:Chaos_in_conservative_discrete-time_systems_subjected_to_parameter_drift}
D.~J{\'a}nosi, T.~T{\'e}l, Chaos in conservative discrete-time systems subjected to parameter drift, Chaos: An Interdisciplinary Journal of Nonlinear Science 31~(3) (2021) 033142.

\bibitem{Negrete_DDC:1996:Area_preserving_nontwist_maps}
D.~del Castillo-Negrete, J.~Greene, P.~Morrison, Area preserving nontwist maps: periodic orbits and transition to chaos, Physica D: Nonlinear Phenomena 91~(1-2) (1996) 1--23.

\bibitem{Negrete_DDC:1997:Renormalization_and_transition_to_chaos_in_area_preserving_nontwist_maps}
D.~del Castillo-Negrete, J.~Greene, P.~Morrison, Renormalization and transition to chaos in area preserving nontwist maps, Physica D: Nonlinear Phenomena 100~(3) (1997) 311--329.

\bibitem{Negrete_DDC:1992:Chaotic_Transport_By_Rosby_Waves_In_Shear_Flow}
D.~del Castillo-Negrete, P.~Morrison, Chaotic transport by {R}ossby waves in shear flow, Physics of Fluids A: Fluid Dynamics 5~(4) (1993) 948--965.

\bibitem{Meiss92}
J.~D. Meiss, Symplectic maps, variational principles, and transport, Rev. Mod. Phys. 64 (1992) 795--848.

\bibitem{Shinohara_S:1998:Indicators_of_reconnection_processes_and_transition_to_global_chaos_in_nontwist_maps}
S.~Shinohara, Y.~Aizawa, {Indicators of reconnection processes and transition to global chaos in nontwist maps}, Progress of Theoretical Physics 100~(2) (1998) 219--233.

\bibitem{Szezech_JD:2009:Transport_properties_in_nontwist_area-preserving_maps}
J.~Szezech~Jr, I.~Caldas, S.~Lopes, R.~Viana, P.~Morrison, Transport properties in nontwist area-preserving maps, Chaos: An Interdisciplinary Journal of Nonlinear Science 19~(4) (2009) 043108.

\bibitem{Szezech_JD:2012:Effective_transport_barriers_in_nontwist_systems}
J.~Szezech~Jr, I.~L. Caldas, S.~R. Lopes, P.~Morrison, R.~L. Viana, Effective transport barriers in nontwist systems, Physical Review E 86~(3) (2012) 036206.

\bibitem{Janosi_D:2024:overview}
D.~Jánosi, T.~Tél, Overview of the advances in understanding chaos in low-dimensional dynamical systems subjected to parameter drift: Parallel dynamical evolutions and ”climate change” in simple systems, Physics Reports 1092 (2024) 1--64.

\bibitem{wurm2004reconnection}
A.~Wurm, A.~Apte, P.~Morrison, On reconnection phenomena in the standard nontwist map, Brazilian journal of physics 34 (2004) 1700--1706.

\bibitem{Chirikov:1979}
B.~V. {Chirikov}, {A universal instability of many-dimensional oscillator systems}, Physics Reports 52~(5) (1979) 263--379.

\end{thebibliography}
\end{document}